\documentclass[12pt,preprint1]{aastex}
\usepackage{color}
\usepackage{amsmath}
\usepackage{hyperref}

\def\be{\begin{equation}}
\def\ee{\end{equation}}

\def\ltsima{$\; \buildrel < \over \sim \;$}
\def\lsim{\lower.5ex\hbox{\ltsima}}
\def\gtsima{$\; \buildrel > \over \sim \;$}
\def\gsim{\lower.5ex\hbox{\gtsima}}

\def\apj{Ap.\ J.}
\def\apjl{Ap.\ J.\ Lett.}
\def\apjs{Ap.\ J.\ Suppl.}
\def\mnras{Mon.\ Not.\ Roy.\ Astro.\ Soc.}
\def\aap{Astron.\ Astrophys.}

\title{Radio-X-ray synergy to discover and study {\it jetted} tidal disruption events}
\author{I. Donnarumma\altaffilmark{1}, E.~M. Rossi\altaffilmark{2}}
\affil{$^{1}$ INAF-IAPS, Via Fosso del Cavaliere 100, 00133, Rome, Italy}
\affil{$^{2}$ Leiden Observatory, Leiden University, P.O. Box 9513, 2300 RA , Leiden, The Netherlands}

\email{immacolata.donnarumma@iaps.inaf.it}

\begin{abstract}
Observational consequences of tidal disruption of stars (TDEs) by supermassive black holes (SMBHs)
can enable us to discover quiescent SMBHs,
constrain their mass function, study formation and evolution of transient accretion disks and jet
formation. A couple of jetted TDEs have been recently claimed in hard X-rays, challenging  jet
models, previously applied to $\gamma$-ray bursts and active galactic nuclei. It is therefore of
paramount importance to increase the current sample.
In this paper, we find that the best strategy
is not to use up-coming X-ray instruments alone, which will yield between several (e-Rosita) and a couple
of hundreds (Einstein Probe) events per year below redshift one. We rather claim that a more efficient 
TDE hunter will be the Square Kilometer Array (SKA) operating {\it in survey mode} at 1.4 GHz. It may detect up to several hundreds of events per year below $z \sim 2.5$ with a peak rate of a few tens per year at $z\approx 0.5$. 
 Therefore, even if the jet production efficiency is {\it not } $100\%$ as assumed here, the predicted rates should be large enough to allow 
  for statistical studies. The characteristic TDE decay of $t^{-5/3}$, however, is not seen in radio, whose flux is quite featureless.
  {\it Identification} therefore requires localization and prompt repointing by higher energy instruments.
If radio candidates would be repointed within a day by future X-ray observatories (e.g. Athena and LOFT-like missions), 
it will be possible to detect up to $\approx 400$ X-ray counterparts, almost up to redshift $2$. The shortcome 
is that only for redshift below $\approx 0.4$ the trigger times will be less than 10 days  from the explosion. In this regard the X-ray 
surveys are better suited to probe the beginning of the flare, and are therefore complementary to SKA. 
\end{abstract}

\begin{document}
\label{firstpage}
\bibliographystyle{apj}

\maketitle


\section{Introduction}
\label{sec:intro}

Since the late 70s it has been suggested that stars torn apart by the 
gravitational field of a
supermassive black hole (SMBH) may be observed as flares from Earth 
\citep{hills,fr76,rees,phinney}.
 These are
called tidal disruption events (TDEs). These flares would be caused by sudden accretion of the
star debris, which would feed the SMBH at an ever decreasing rate, $\dot{M} \propto
t^{-5/3}$. This theoretical expectation is for a complete disruption of a star in parabolic orbit,
after at least several days from the peak \citep[e.g.][]{lodato09,hayasaki,guillochon}, and it is
expected to be independent on the ratio of pericenter to tidal radius
\citep{sari2010,stone2013}.

The detection and study of these flares can deliver important astrophysical information. On the
one hand, they allow us to detect otherwise quiescent SMBHs and estimate their masses. This would
inform theory of galaxy-SMBH cosmological co-evolution. On the other, they constitute a unique
opportunity to study the -- highly theoretically uncertain -- formation of an accretion disc and its
continuous transition through different accretion states. As the accretion rate decreases, we can
in principle observe a disc which transits from an initial super-Eddington phase, lasting several
months, passing through a \emph{slim} and later a \emph{thin} disc regime, and ending its
life, years later, in a radiative \emph{inefficient} state.
The super-Eddington phase --which occurs only for SMBH masses $ \lsim 10^7$ M$_{\odot}$ --- is
highly uncertain, but it may be associated with a copious radiative driven wind \citep{rossi09},
which thermally emits $\sim 10^{41}-10^{43}$ erg s$^{-1}$, mainly at optical frequencies
\citep{strubbe2009,lodatorossi}. The disc luminosity ($\sim 10^{45}-10^{46}$ erg s$^{-1}$),
instead, peaks in  far-UV/soft X-rays \citep{lodatorossi}.  Of paramount theoretical importance
would also be the possibility to investigate the formation and evolution of an associated jet,
powered by this sudden accretion.
There is no specific theory for the jet emission from TDEs. Astronomers mainly assume a
phenomenological description \citep[e.g.][]{vanvelzen11,cannizzo} or borrow theory developed 
for blazars and/or
$\gamma$-ray bursts \citep[e.g.][]{metzger12,sasha}. In general, non-thermal emission in X-rays and
radio is the jet signature.

Handful of candidate TDEs ($\sim 10$) have been detected so far, particularly in ROSAT all sky
survey \citep{komossa, donley}, in GALEX Deep Imaging Survey \citep{gezari2009,gezari2012,scampana} 
and in SDSS \citep{vanvelzen11a}.
 These ``soft" events are believed to be associated with the disc and wind thermal emission. The
presence of a bright optical flare in the initial super-Eddington months makes optical surveys a
useful tool for discovery. Significant advances in optical transient surveys are expected to be
achieved by the Panoramic  Survey  Telescope and Response System (Pan-STARRS) and the Large
Synoptic Survey Telescope (LSST).
Two candidates have been claimed in Pan-STARRS data  \citep{gezari2012, chornock14},
 three in PTF data \citep{arcavi} and one in ASAS-SN \citep{holoien}, 
but the total number expected seems to be much higher.  
For example in the $3\pi$ Survey, claims in literature range from 200  to $\sim 1557$,
while in the medium deep survey there is more consensus that $\sim 15-20$ 
should be found \citep{strubbe2009, vanvelzen11a}.
{\em Thousands} of candidates could be, instead, detected by LSST, with its 6-band ($0.3-1.1$ micron)
wide-field deep astronomical survey of over 20000 square degrees of the southern sky, using an
$8.4$-meter ground-based telescope \citep{strubbe2009, vanvelzen11a}. However, these estimates are 
probably upper limits, because
galactic nuclei can heavily absorb optical light.

More recently, two candidates TDEs were  triggered in the hard X-ray band
by the BAT instrument on board of Swift \citep{bloom, burrows, cenko}.
A multi-frequency follow-up from radio to $\gamma$-rays revealed a new class of TDEs, where we are
likely observing the non-thermal emission from a  relativistic jet. The jet emission is responsible for the
hard X-ray spectrum (with power-law slope $\beta \sim 1.7$) and the increasing radio activity
\citep{levan2011}, detected a few days after the trigger.

Given the lack of statistics and of a solid theoretical framework for the non-thermal emission, we
will take the best studied of these two events, Swift J1644+57 (Sw J1644 in short), as a prototype for
the study presented in this paper, where we investigate the detection capability of both SKA and
future X-ray observatories.

Sw J1644 was hosted by a star forming galaxy at $z=0.354$ and in positional coincidence 
with its center \citep{zauderer}.
  Its X-ray peak luminosity  $\sim 3 \times 10^{48}$ erg s$^{-1}$ was reached after a couple 
of days from the trigger, and it persisted
at the level of $>10^{45}$ erg s$^{-1}$ for about 1 year. During its decay, the X-ray emission was
approximately described by a $t^{-5/3}$ temporal law, the same as that expected for the fallback of
stellar debris (see Figure \ref{fig:x_j1644}). After $\sim 500$ day from the trigger, 
the X-ray flux declined by two 
orders of magnitude and it has been
associated with a shut off of the relativistic jet \citep{zauderer13}.
 The modelling of the X-ray luminosity suggests 
that Sw J1644 is associated with a light supermassive black hole $\lsim 10^{7} M_{\sun}$
\cite[e.g.][]{burrows,cannizzo}.

Variability at optical wavelengths within the host was not detected, while transient emission was
seen in infrared, becoming stronger at longer wavelengths, especially at millimeter and radio
wavelengths.  Radio (1.4. and 4.8 GHz) observations from Westerbork Synthesis Radio Telescope (WSRT)
showed a bright source. EVLA
observations of the radio transient coincident with the host galaxy were reported, 
providing an estimate of the
bulk Lorentz factor $\Gamma \sim 2$ of the outflow  \citep{zauderer}.
The radio lightcurve displays a rebrightening  starting one
month after the trigger \citep{berger12, zauderer13}. The emission peaks around several months,
followed by a decline.
Radio observations stop at 600 days after the trigger \citep{zauderer13}. 
The radio behavior is not compatible with the blast wave
model borrowed from $\gamma$-ray bursts by \cite{metzger12}, and indicate 
a more complicated jet structure, like
perhaps in the magnetically arrested model proposed by \cite{sasha}.  Snapshot rates of jetted TDEs
in radio band have been computed for the first time by Van Velzen et al. (2011).
Differently from their work, we adopt here a different modelling for the radio lightcurve and a more
detailed one for the black hole mass function, which includes the redshift dependence. We also account for a stellar mass function. 
We broaden up our investigation to include X-ray detection and follow-ups.

Finally, a 200-s quasi-periodic oscillation (QPO) was detected by both Suzaku and XMM, $\sim 10$ and
19 days after the Swift/BAT trigger, respectively \citep{reis}. QPOs are regularly detected in
stellar mass BHs, but there is no firm physical interpretation of these phenomena. However, most
models strongly link the origin of high-frequency QPOs with orbits or resonances in the inner
accretion disk close to the BH. This may cause variable energy injection into the jet, which
consequently results in variability in the X-ray emission. This interpretation led  to
estimate a BH mass between $ 5 \times 10^{5} M_{\odot}$ and $ 5 \times 10^{6} M_{\odot}$ \citep{reis}.

In this paper, we predict the detection rate of {\it jetted} TDEs 
considering current and future radio surveys 
(NVSS + FIRST, VLT Stripe 82, ASKAP, VLASS and SKA) and
 X-ray instruments (Swift, eRosita, Einstein Probe, Athena, LOFT).
 In addition, we discuss the ability of these 
 instruments to constrain important physical parameters.

The paper is organized as follows.
In \S 2, we take Swift J1644 as a prototype and we describe our phenomenological 
model for X-ray and Radio emissions.
In \S 3,  we discuss the black hole distribution functions used in this paper. 
In \S 4, we present our
 Monte Carlo calculations. Our rates for current and future surveys
are presented in \S 5. A summary and implications  of our results can be found in \S 6. 
Finally, we draw our conclusions in \S 7.

Throughout this paper we use the following cosmological parameters:
 $\Omega_{\rm M}= 0.25$, $\Omega_{\rm \lambda}= 0.75$ 
and $H_{\rm 0}=70~ {\rm km/s/Mpc}$.

\section{Modelling the Lightcurve}

A tidal disruption event of a star by a SMBH causes a transient accretion disc to form,
whose accretion rate is set by the rate at which the stellar debris falls back to the black hole
under its gravitational pull. How matter circularizes to form a disc and whether this process is accompanied
by outflows and their characteristics are subject to intense investigations, as mentioned above.
From phenomenology and theory, we know that in the presence of an accretion disc and some
ordered magnetic field, matter and energy outflows in form of (relativistic) jets are produced.
In the absence of fully consistent simulations of jet production by a tidal disruption event,
we use below a simplified description for the jet energy content as a function of time. This is
partially supported by analytical and numerical calculations (see references above) and partly by
 the observed features of the X-ray emission of Sw J1644.
In particular, its temporal decay ($\sim 
t^{-5/3}$) suggests that at least in this optically thin regime, the X-ray 
luminosity scales as the accretion rate. As a consequence, it 
supports a scenario in which the star was 
completely tidally
disrupted, since partial disruption would lead to a shallower decay of the 
fallback rate \citep{guillochon}. Moreover, a 
partial disruption is difficult to reconcile with a long lasting super Eddington 
accretion phase, which may be needed to power the jet for its total duration of $\sim 500$ days. 
Finally, the modelling of the X-ray luminosity suggests 
that Sw J1644 is the consequence of a disruption of a roughly one solar mass star by a light 
supermassive black 
hole $\lsim 10^{7} M_{\sun}$
\cite[e.g.][]{burrows,cannizzo}.

\subsection{Jet kinetic power}
We work in the framework of two identical jets, with $\theta_{\rm j} < 1/\Gamma$.
The total energy injected in the two jets is $L_{\rm j}= \epsilon_{\rm j} \dot{M}_{\rm fb} c^2$, where
$\epsilon_{\rm j}$ is the jet production efficiency, which we assume constant in time, and the gas fall back to form a disc occurs at a rate
$\dot{M}_{\rm fb}$. For a complete disruption of a star in parabolic orbit the fallback rate can be approximated by
\begin{equation}
 \dot{M}_{\rm fb}(\bar{\tau}) = \dot{M}_{\rm p} \left(\frac{t_{\rm min}+\bar{\tau}}{t_{\rm 
min}}\right)^{-5/3},
 \label{eq:mdot}
\end{equation}
\citep{rees,phinney}. The lag time ``$\bar{\tau}$'' is the time from the beginning of the 
debris accretion, that 
roughly happens after a time 
$$t_{\rm min} \approx 41 ~M_6^{1/2}~{m_{*,1}^{1/2}~\rm day},$$
from the star disruption, in the galaxy rest frame. More precisely, $t_{\rm min}$ is the 
minimum 
time it takes the most bound debris to come back to pericenter after the star has been torn apart. 
Here and in the following, $M_6$ is the BH mass in units of $10^6 M_{\sun}$ and $ m_{*,1}$ the 
mass of the disrupted star in units of $1 M_{\sun}$. The peak of the accretion rate\footnote{In 
the formula used in this paper, we 
assume the standard linear relation between mass and radius of the star. See eq.6 in 
\citep{lodatorossi}.} is 
 quite intuitively the mass of the star divided by the characteristic timescale, $\dot{M}_{\rm p} 
\approx (1/3)\, 
m_*/t_{\rm min} \approx 1.9 \times 10^{26} M_{6}^{-1/2} m_{*,1}^{1/2}$ g s$^{-1}$.
 In our description, the jet is launched at the onset of accretion ($\bar{\tau}=0$), as there are 
no strong theoretical reasons why it should be delayed. The temporal evolution of the jet energy 
is thus

\begin{equation}
L_{\rm j}(\bar{\tau}) =  L_{\rm j,p} \left(\frac{t_{\rm min}+\bar{\tau}}{t_{\rm min}}\right)^{-5/3},
\label{eq:Lj}
\end{equation}
where
\begin{equation}
L_{\rm j,p} =  \epsilon_{\rm j} \dot{M}_{\rm p} c^2 \approx
1.7 \times 10^{45} ~{\rm erg ~s}^{-1} \left(\frac{\epsilon_{\rm j}}{0.01}\right) M_6^{-1/2} m_{*,1}^{1/2}.
\label{eq:Ljp}
\end{equation}
Note that the larger the black hole mass, the lower the peak luminosity, because the characteristic 
timescale increases. Viceversa, the jet luminosity decreases with $m_{*}$.

\subsection{X-ray}
\label{sec:X-ray_modelling}

\begin{figure}
\includegraphics[width=\columnwidth]{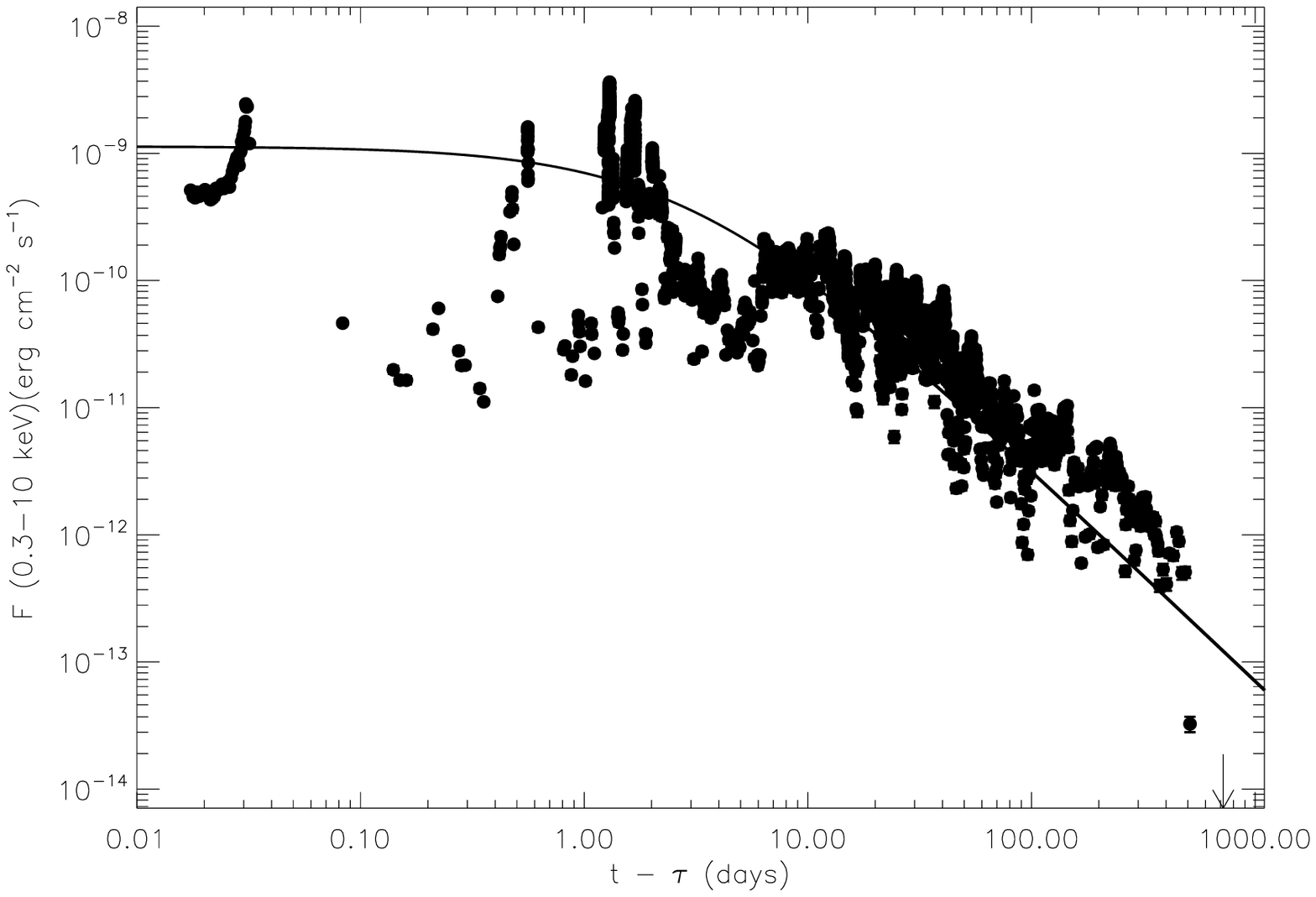}
\caption{The X-ray (0.3-10 keV) lightcurve of J1644 as a function of time from the X-ray trigger: data (absorbed flux, circle marks taken from 
\href{http://www.swift.ac.uk/xrt_curves/}{Publicy available XRT lightcurves}) versus our modelling (solid line). After a few days, the temporal decay approaches $t^{-5/3}$.}
\label{fig:x_j1644}
\end{figure}

The {\it unabsorbed} 1-10 keV lightcurve of Sw J1644 is shown in Fig.\ref{fig:x_j1644} 
(black cirles).
Activity was already detected by BAT $\approx 3$ days 
before the BAT ``official'' trigger and the beginning of XRT observations \citep{burrows}. 
Therefore there is an indication that the trigger (i.e. when the first photon was detected by XRT) 
happened approximately $\tau\approx 3$ day {\it after} the actual disc and jet 
formation. The observed time interval $\tau$ is related to the rest frame 
analogous quantity by $\tau = \bar{\tau} (1+z)$ and in this case $\bar{\tau} 
\approx 2$ day. 
Accounting for this delay, the general behaviour of the X-ray 
lightcurve as a function of time 
$\Delta {\rm t}$ since the trigger ($\Delta {\rm t}=0$) can be reproduced by
\begin{equation}
L_{\rm x,iso}(\Delta t) \approx 1.5 \times 10^{48} ~{\rm erg ~s^{-1}}~ \left(\frac{\tau+\Delta {\rm 
t}}{\tau}\right)^{-5/3},
\label{eq:Xfit}
\end{equation}
(Fig. \ref{fig:x_j1644}, solid line). Specifically, $L_{\rm x,iso}$ is an isotropic equivalent 
luminosity, computed from the X-ray flux. Note that here $\tau=3$ is a fixed time delay, 
unlike $\bar{\tau}$ in eqs.\ref{eq:mdot} and \ref{eq:Lj}. Superimposed to this baseline trend, 
there is a 
complex structure of flares and dips where the flux oscillates within two orders of 
magnitude in the first ten days of observations. It is clear that eq.\ref{eq:Xfit} does not capture
this large variability, possibly associated with jet 
precession and nutation \citep{saxton,stone2012}. But in absence of a compelling theory for these sudden 
X-ray variations, we prefer to reproduce the upper part of the envelope that contains the initial 
variability, since the BAT instrument was triggered by one of the peaks in the lightcurve. We 
will discuss later how this choice affects our X-ray TDE rate estimates.

The Swift/XRT (0.3-10 keV) spectrum of Sw J1644+57 is well described by an absorbed power-law 
with a photon index 
$\beta \approx 1.7-1.8$ and $N_{\rm H}
\approx 2 \times 10^{22}$ cm$^{-2}$ \citep{burrows}.
The observed BAT spectrum at early times
{\it and} its count rate later on (up to the beginning of June) are consistent with an
 extrapolation at higher 
energies of the XRT spectrum  \citep{burrows}.
This suggests that we are observing the same component in both soft and hard X-ray bands. 
The average spectrum is consistently hard ($1.4 \lsim \beta \lsim 1.7$) during the whole emission, 
although a spectral softening is observed during the short dips in the initial variable phase 
\citep{saxton}.
The radiation efficiency  in 1-10 keV band (i.e. the fraction of the total luminosity 
emitted in that band) is $\epsilon_{\rm x} \approx 0.20$ 
\citep{burrows}.
With this last information, we can calculate
the associated {\em jet kinetic luminosity} from the observed light curve, once we assume a 
jet opening angle $\theta_{\rm j}$ 
and a Doppler factor $\delta$, 
$$ L_{\rm j} = L_{\rm x,iso} (1-\cos{\theta_{\rm j}})/(\epsilon_{\rm x} 
\delta^2).$$

With the highest  probability, our line of sight is at an angle $\sim 
\Gamma^{-1}$ (i.e. the inverse of the Lorentz factor $\Gamma$) that grazes 
the 
relativistic beam, and  $\delta \approx \Gamma$.
The fact that there are no sharp breaks in the lightcurve may indicate that the whole emitting area was visible, i.e. $\Gamma^{-1}> \theta_{\rm j}$. Therefore, we further assume a jet opening angle of a similar size of the relativistic beaming, say $\theta_j \approx \Gamma^{-1}/2$, and
we get a jet power at the trigger time ($\Delta t=0$) of $L_{\rm j}(\bar{\tau}) \approx 1.5 \times 
10^{45} 
~{\rm erg 
~s^{-1}} (\Gamma/5)^{-4}$. Since $L_{\rm j}(\bar{\tau}) = 
\epsilon_{\rm j}\dot{M}_{\rm fb}(\bar{\tau})c^{2}$, it turns out that to have an efficiency 
$\epsilon_{\rm 
j}$ greater than $1\%$ requires $m_* \le 1 M_{\odot}$, for $\Gamma \le 5$. In 
particular,
$m_* =1 M_{\odot}$ gives efficiency between roughly $1\%$ and $37\%$ for $2 \le \Gamma \le 5$, that 
are in agreement with numerical simulations of jets from highly super-Eddington accretion discs 
\cite{sadowski}. Lower mass stars would give a higher efficiency range.
We therefore assume in the following that Sw J1644 is the result of the disruption of a solar mass 
star. However, it is clear that this is just a tentative, though reasonable, choice, since the 
stellar mass cannot in fact be univocally determined, unless we can actually measure $\theta_j$.

Assuming Sw J1644 as a prototype, we can adopt a general description of the X-ray lightcurve 
in the 1-10 keV band, 
when we catch the flare after a time $\tau$ from the beginning of the event,
\begin{equation}
L_{\rm x,iso}(\Delta t) = L_{\rm x,t}~\left(\frac{\tau+\Delta {\rm t}}{\tau}\right)^{-5/3}.
\label{eq:lx}
\end{equation}
The (isotropic equivalent) luminosity $L_{\rm x,t}$ at the time of the trigger ($\Delta {\rm t} =0$) is
$L_{\rm x,t} = L_{\rm j}(\bar{\tau}) \epsilon_{\rm x}\, \delta^2/(1-\cos{\theta_{\rm j}}) \simeq 
 L_{\rm j}(\bar{\tau})\epsilon_{\rm x} 2 \left(\Gamma/\theta_{\rm j}\right)^2$, which can be written 
more 
explicitly as
\begin{equation}
\begin{split}
L_{\rm x,t} & \approx 1.63 \times 10^{48} ~{\rm erg ~s^{-1}} M_6^{-1/2} m_{*,1} 
^{1/2} \left(\frac{\epsilon_{\rm 
x}(z)}{0.2}\right) \\
 & \times \left(\frac{t_{\rm min}+ \bar{\tau}}{t_{\rm min}}\right)^{-5/3} ,
 \label{eq:lx_trigger}
\end{split}
\end{equation}
 where $\bar{\tau} = \tau/(1+z)$ and the radiation efficiency $\epsilon_{\rm x}(z)$ 
varies because of the spectral shifting 
with redshift,
\begin{equation}
\label{eq:kcorr}
\epsilon_{\rm x}(z) = 0.20 \frac{(E_2 (1+z))^{-\beta +2} - (E_1 (1+z))^{-\beta+2}}{(E_2 (1+z_{\rm 
sw}))^{-\beta +2} - (E_1 (1+ z_{\rm sw}))^{-\beta+2}},
\end{equation}
where we assume $\beta=1.8$, $E_1=1$ keV, $E_2=10$ keV and z$_{\rm sw}=0.35$.
 We note that this correction is in the source rest-frame and applies to unabsorbed fluxes.

In eq.\ref{eq:lx_trigger}, we set $\epsilon_j \Gamma^2/\theta_{\rm j}^2 \approx 23.9$.
Indeed, any combination of these quantities that gives a factor $\approx 24$, allows us to 
reproduce the Sw J1644 X-ray luminosity at the trigger time. The 
degeneration should then be lifted,
when we need to choose a Lorentz factor to compute the TDE rates.
From the X-ray luminosity, the flux is easily computed, $$F_{\nu}= \frac{L_{\rm 
x,iso}}{4 \pi D^2},$$ where $D$ is the
 luminosity distance.

\subsection{Radio Lightcurve}
\label{radio}
In this section, we first reproduce the lightcurve at 1.4 GHz of Sw J1644 and then we 
generalize it to events at different
 redshifts and with different
stellar and black hole masses.

The radio emission is synchrotron emission and the low energy spectrum can be described with the 
following broken power-law
\begin{equation}
\begin{split}
 F_{\nu} &= F_{\nu}(\nu_{\rm a}) \left[\left(\frac{\nu}{\nu_{\rm a}}\right)^{-2 s_1 } +\left(\frac{\nu}{\nu_{\rm a}}\right)^{-s_1/3}\right]^{-1/s_1} \\
 & \times \left[1 +\frac{\nu}{\nu_{\rm m}}^{s_2}\right]^{-1/s_2},
 \label{eq:spectrum_radio}
\end{split}
\end{equation}
\citep{granot} where $\nu_{\rm a} < \nu_{\rm m}$ are respectively the absorption and peak frequency,
$s_1,s_2$ are smoothing factors and the electron power-law index has been assumed to be $2.5$.

\cite*{berger12} measure the flux $F_{\nu}(\nu_{\rm a})\equiv F_{\nu,sw}(\nu_{\rm a,sw})$, 
and characteristic frequencies $\nu_{\rm a}\equiv \nu_{\rm a,sw}$ and
$\nu_{\rm m} \equiv \nu_{\rm m,sw}$, in several snapshots that cover the evolution
of the lightcurve up to $\sim 220$ days after the trigger. Later, \cite{zauderer13} extended
 the period of the radio monitoring up to $\sim 600$ days.
The first observation is at $\sim 5$ days after the detection in the X-ray band.
Therefore, the radio emission is observed after a delay $\tau \simeq 8$ days from the
 intrinsic beginning of the event. 
Finally, note that the radio data monitoring occurs up to $\sim 600$ days \citep{berger12, zauderer13}, while the X-ray emission
 has been observed up to $\sim 500$ days. This mismatch, however, is not a problem, since we are interested in
 modelling the lightcurves only up to one year after the explosion, when is already too dim to be detected by an X-ray survey
  in most cases.

\begin{figure}
\includegraphics[width=\columnwidth]{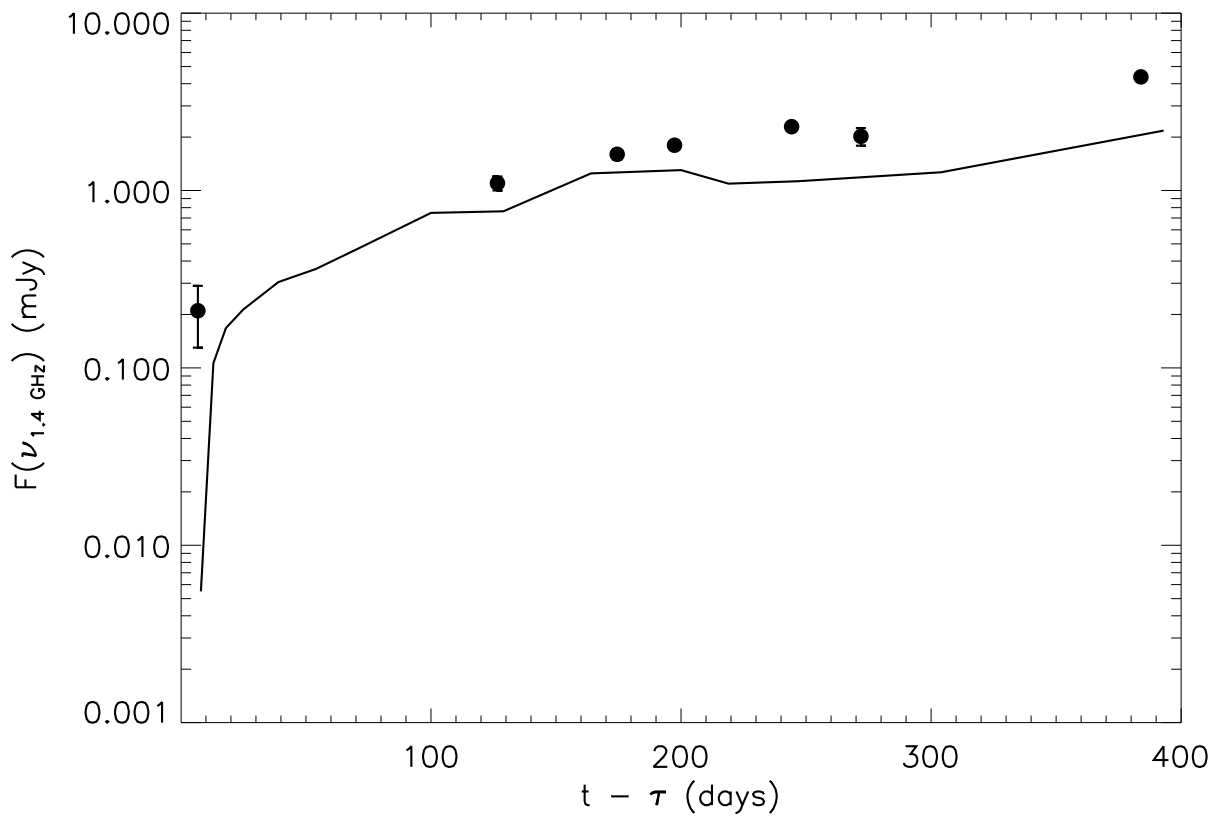}
\caption{The radio (1.4 GHz) light curve of SwJ 1644 as a function of time from the radio trigger (5 days after the X-ray first detection): data (circle marks) from \citep{berger12,zauderer13} versus 
our modelling (solid line). While our modelling well reproduce higher radio frequencies lightcurves (see  fig. 1 in  \citep{berger12}), it slightly 
underpredicts the 1.4 GHz one. In this respect our flux modelling is conservative.}
\label{fig:radio_j1644}
\end{figure}

Using the available data and eq.\ref{eq:spectrum_radio}, we can therefore model 
the temporal evolution of the flux at any radio frequency.
In Figure \ref{fig:radio_j1644}, we show the lightcurve of Sw J1644 at 1.4 GHz, and 
its comparison with data. A smooth temporal behavior 
has been obtained by linearly interpolating the flux between data points.

We now need to generalize our prototypical lightcurve to a generic TDE. 
The main uncertainty is how the flux scales with black hole and stellar
masses. A first possibility is to describe the jet evolution with a Blandford Mckee 
(thereafter ``BM model") solution, usually adopted for $\gamma$-ray burst
afterglows  \cite[e.g.][]{metzger12,berger12}.
Frequencies below 5 GHz are in the self-absorbed part of the synchrotron spectrum, 
for the whole observed duration of the event
\cite[see fig.3 in][]{berger12}. The observed specific luminosity in this regime ($\nu <\nu_{\rm a} < \nu_{\rm m}$)
is given by the Raleigh Jeans part of the Black Body spectrum $B(\nu/\delta)\delta^3 \propto k_{\rm b} T (\nu/\delta)^2 \delta^3$ (see eq. \ref{eq:spectrum_radio}),
with a kinetic temperature  given by $3 k_{\rm b} T = m_{\rm e} c^2 \gamma_{\rm min}$,
where the minimum Lorentz factor for the shocked accelerated electrons is $\gamma_{\rm min} \propto \Gamma$.
Therefore the specific radio luminosity is
\begin{equation}
L_{\nu} \propto B(\nu/\delta) \delta^3 (r \theta_j)^2 \propto (r \Gamma)^2,
\label{eq:Lvr}
\end{equation}
where $(r \theta_{\rm j})^2$ is the emitting area, and we are assuming $\Gamma^{-1} \gsim \theta_{\rm j}$.
In the blast wave modelling of J1644, the external 
medium swept up by the jet is 
better described by a power-law density decay that goes as
$r^{-2}$, rather than a constant density environment \citep{zauderer}. 
This implies $\Gamma \propto E_{\rm j}^{1/4}$ and $r \propto  E_{\rm j}^{1/2}$,
where $E_{\rm j} \approx L_{\rm j,p} t_{\rm min} \propto m_*$ is the total jet energy. Therefore 
eq. \ref{eq:Lvr} becomes,
$L_{\nu} \propto  (L_{\rm j,p} t_{\rm min})^{3/2} \propto m_*^{3/2}$, where 
there is no dependence on the black hole mass, but only on the stellar mass.

The simple blast wave solution, however, does not describe the whole evolution 
of the radio spectrum \citep{berger12}. Therefore, we also consider a simpler approach. 
In line with our treatment of the X-ray flux, we may assume that the radio luminosity is 
proportional to the jet peak luminosity
$ L_{\nu} \propto L_{\rm j,p} \propto M^{-1/2} m_*^{1/2}$, rather than to its total energy,
(see the X-ray analogous, eq.\ref{eq:lx_trigger}, which bears the same mass 
dependencies). As an extra motivation, this prescription may be justified 
in the context of the ``magnetically arrested" jet model \cite[e.g.][]{narayan03}.
We will call this prescription ``the Mass Dependent Luminosity"  model (thereafter MDL model).

The scaling of the peak flux for sources at different redshift, with different black hole and 
stellar masses (but at the same observed time $\tau$ from the beginning of the event) would be
\begin{equation}
F_{\nu}(\nu_{\rm a},\tau)=    F_{\nu,sw}(\nu_{\rm a,sw},\tau_{\rm sw}) 
\times 
m_{*,1}^{3/2} 
\left(\frac{1+z}{1+z_{\rm sw}}\right) \left(\frac{D_{\rm sw}}{D} \right)^2,
\label{eq:Lvradio}
\end{equation}
for the BM solution and
\begin{equation}
F_{\nu}(\nu_{\rm a},\tau)= F_{\nu,sw}(\nu_{\rm a,sw}, \tau_{\rm sw}) \times M_6^{-1/2} 
{m_{*,1}^{1/2}}\, \left(\frac{1+z}{1+z_{\rm sw}}\right) \left(\frac{D_{\rm sw}}{D} \right)^2,
\label{eq:Lvradio_Mbh}
\end{equation}
for our second approach. The equivalent delay at which we need to calculate the flux of Sw J1644 is 
$\tau_{\rm ws} \equiv \tau \frac{1+z_{\rm sw}}{(1+z)}$.

The characteristic frequencies need to be 
redshifted\footnote{Formally, 
one would need 
to consider the transformation due to different Doppler factors between jets. However, 
we here assume that all 
jets have approximately the same Lorentz factor $\Gamma$ and viewing angle 
of nearly $\theta_{\rm o} \approx 1/\Gamma$. The latter is because the viewing angle probability 
($\propto \theta_{\rm o}$, between 
$0<\theta_{\rm o}< \Gamma^{-1}$) is the highest at $\Gamma^{-1}$.} 
according to
$$\nu_{\rm a}(\tau)=\nu_{\rm a,sw}(\tau_{\rm sw}) \left(\frac{1+z}{1+z_{\rm sw}}\right)^{-1},$$ and $$\nu_{\rm m}(\tau)=\nu_{\rm m,sw}(\tau_{\rm sw}) \left(\frac{1+z}{1+z_{\rm sw}}\right)^{-1}.$$
In all cases, the flux $F_{\nu,sw}(\nu_{\rm a,sw}, \tau_{\rm sw})$ and the characteristic frequencies 
at any time $\tau_{\rm sw}$ are obtained 
by linearly interpolating the available data. For $\tau_{\rm sw} < 8$ day we extrapolate the radio 
light curve to earlier epochs.

\section{Black hole mass functions}
\label{sec:BH_mas_functions}

\begin{figure}[h!]
\begin{center}
\epsscale{1.7}
\plottwo{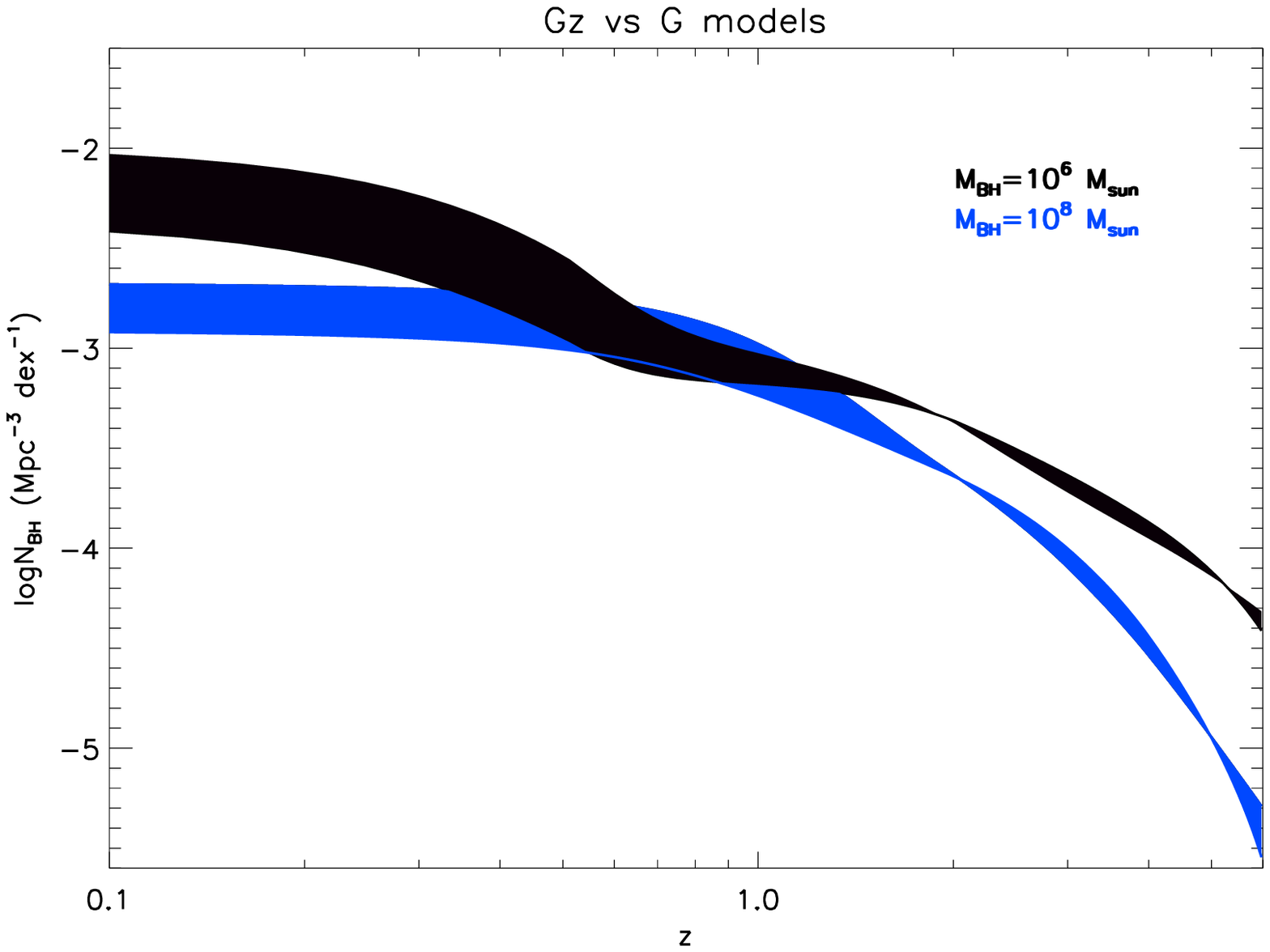}{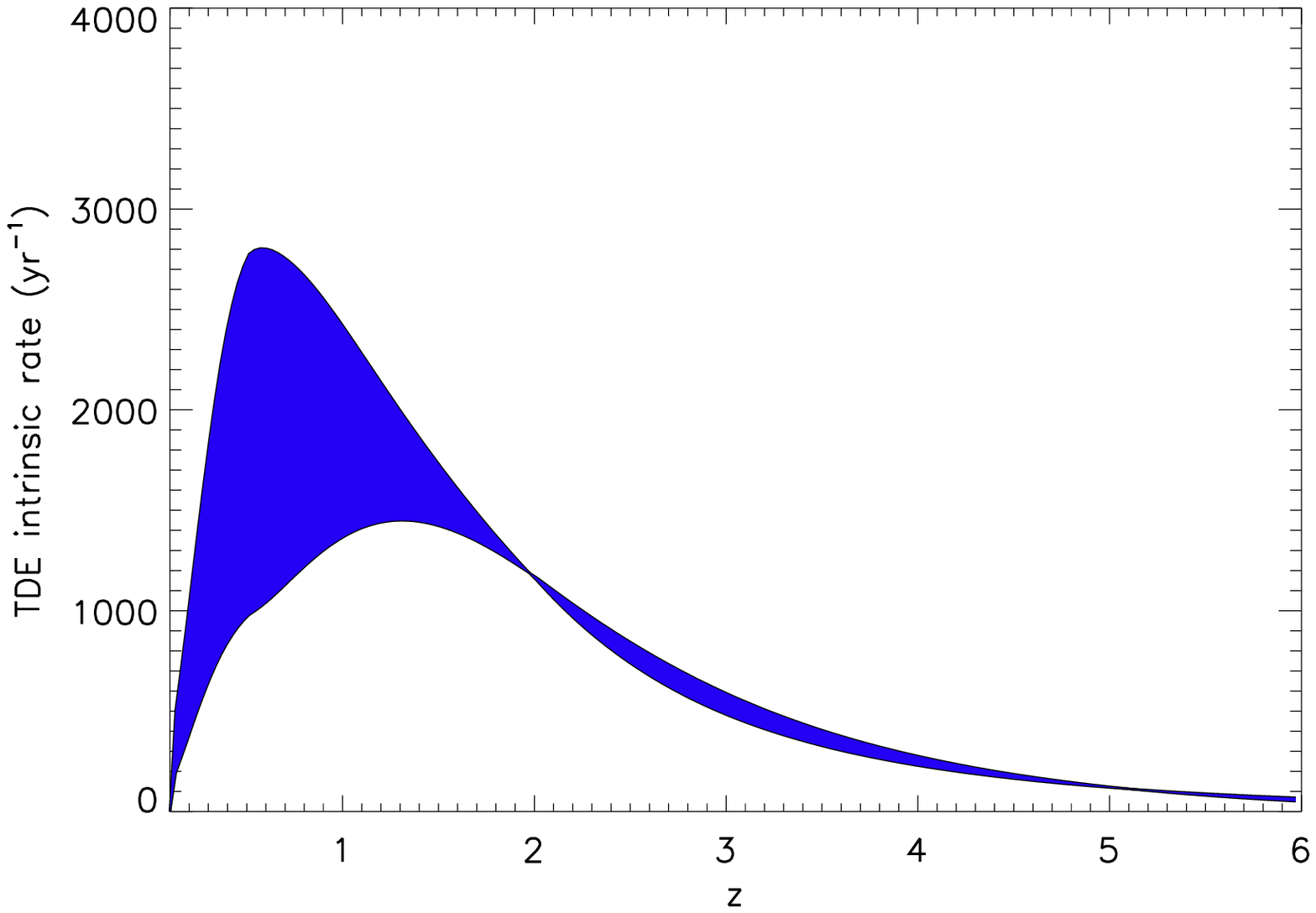}
\caption{Upper panel: BH number density as a function of $z$ for $10^{6}$ (black shaded area) and $10^{8}$ M$_{\odot}$ (blue shaded area).
 Lower panel: intrinsic rate of TDEs as a function of redshift. A rate of $10^{-5}$ yr$^{-1}$ per galaxy is assumed. Most of the events are expected below $z\sim 2$.}
\label{fig:rate_BH}
\end{center}
\end{figure}

The mass distribution of black holes as a function of redshift is an essential ingredient to
 calculate TDE rates.
Since black holes grow mainly by efficient accretion \citep{soltan}, one can calculate these functions using the mass continuity 
equation, given a radiation efficiency and distribution
of Eddington ratios. In this paper, we use the results from \cite{shankar13}. In particular, 
we consider the two 
accretion models which yield the largest and the lowest
black hole comoving number density $\phi(M,z)$, {\em and} are still consistent with the quasar
 bolometric luminosity 
functions and the local black hole mass function
\citep[models labeled $G$ and $G(z)$ in][]{shankar13}. In this way, we can estimate the uncertainty 
due to the black hole mass distribution of our expected TDE rates. In Figure \ref{fig:rate_BH} upper
 panel, we show 
the mass distribution functions and their uncertainty strips
 as a function of redshift,
for $M=10^6 M_{\sun}$ and $M=10^8 M_{\sun}$ black holes.
In Figure \ref{fig:rate_BH}, instead, we show the ``intrinsic'' TDE rate as a function of redshift,
\begin{equation}
 R(z) = \int_{M_{\rm min}}^{M_{\rm max}} \phi(M,z) V(z) N_{\rm tde} dM,
 \label{eq:mf}
\end{equation}
where we denote with $V(z)$ the comoving cosmological volume. $N_{\rm tde} = 10^{-5}$ yr$^{-1}$ is  our
fiducial TDE rate per galaxy: this value is in the range
of theoretical expectations \citep{merritt13} and observational claims 
\citep{donley,gezari2009,vanvelzen11}.

The minimum  black hole mass (here and thereafter in our calculations)
is $M_{\rm min} =10^{6} M_{\sun}$,
 as just a few SMBHs have been observed with a lower mass.\footnote{The recent discovery of TDEs
 in dwarf galaxies ($\lsim 10^{6} \rm M_{\sun})$ \citep{donato,maksyma,maksymb} seems particularly 
promising in overcoming this limit and use TDEs to find lower mass BHs}

\section{Monte Carlo calculations}
\label{sec:MC}
Assuming the X-ray and radio modelling described in \S2, we perform Monte Carlo simulations (MCs)
 to derive the number of jetted TDEs to be detected per year, for given flux limit and sky coverage.

Beside the BH mass, the main ingredients of our MCs are the trigger lag time, $\tau$,  and the mass of the 
disrupted stars, $m_*$. The former is randomly extracted  from a uniform distribution between 
0 and 1 yr\footnote{we do not use longer time lags because any extrapolation beyond the currently available radio data would make our estimates more model-dependent, since there is no hydro-dynamical model that can reproduce the whole radio behavior of J1644.}. The latter follows a Kroupa Initial Mass Function,  \cite[IMF][]{kroupa},
\be
f(m)   \propto \left\{\begin{array}{lll}
m^{-0.3},  & \,\; 0.01 \le m_*\le 0.08, \\
m^{-1.3}, &\,\; 0.08 \le m_*\le 0.5, \\
m^{-2.3}, &\,\; m_* > 0.5.
\label{eq:fv_ae}
\end{array}\right.
\ee
 In fact, for each black hole mass, the minimum stellar mass is set by the requirement that the tidal radius should be greater than the last stable orbit (we assume a non-spinning BH). This requirement implies that $m_{\rm *min} =\max[0.01, 0.045 M_6]$. Note that for $M=10^{8} M_{\odot}$, the minimum mass is  $m_{\rm *min} =4.5 M_{\odot}$. Therefore, events associated with high BH masses are suppressed in numbers by the steepness of the IMF, as only $0.4\%$ of all stars have  $m_* > 8$. However, they are in average brighter, because the average $m_*$ is larger.

In our simulation, we start by considering the intrinsic rate $R(z)$  (eq. \ref{eq:mf})
properly modified by accounting for the relativistic beaming, which results in a reduction by a factor of $ 2 \pi\Gamma^{-2}/(4 
\pi)= 1/(2 \Gamma^2)$: this is the fraction of solid angle subtended by
the emission, when considering a two sided jet. Our
fiducial value for the jet Lorentz factor is $\Gamma=2$, as inferred by
radio observations \citep[$\Gamma \approx 2$, ][]{zauderer13,berger12}.
If the jet decelerates, this value has to be intended as an
average one, over the observation period.
However, we note that this is a geometrical scaling factor and our results may be easily re-scaled by assuming 
different values of the jet bulk Lorentz factor.
In addition,  $R(z)$ is scaled for the fraction of the sky surveyed by the assumed instrument.
In the calculation of $R(z)$ we have adopted both $G$ and $G(z)$ models in order to account for
 the systematic uncertainties 
in the mass function modellings. 
The number of trials in MCs is properly fixed by requiring a high statistics level in each mass and 
redshift bin (typically $\geq 10^{4}$).

\section{Results}
In this section, we first validate separately our emission models for X-ray and radio light curves,
 by comparing our 
predicted rates with current
instruments and survey results. In fact, we find that current data do {\it not } put strong 
constraints on our modelling, 
as we will explain in the following. Future data have instead a greater potential. 
 In the SKA era, we propose that a strategy where radio will be triggering
 X-ray facilities  can allow us not only to detect but also to {\it identify}
and {\it investigate} jetted TDEs in a multi-wavelength fashion. In the following, if not otherwise mentioned, our results are derived adopting $\Gamma=2$. 

\subsection{Comparison with current surveys}
\subsubsection{Hard X-rays}
\label{sec:hardX}

So far, only two jetted TDE candidates have been detected by BAT, implying an observed rate of
$ \sim 0.3$ yr$^{-1}$. 

Since BAT is not operating in survey mode, it is not straightforward to
compare observations with our predictions, i.e. it is difficult to chose sky coverage and 
detection limit, because they are not univocally determined. 
The two TDE candidates were detected in two different modes: Sw J1644 was triggered onboard, while Sw J2058 was discovered  by stacking 4-day integration images \citep{krimm11,cenko}. In both  modes, it is hard to define a survey flux limit, the key ingredient of our MCs. Indeed, Swift has  {\it over} 500 onboard trigger criteria in 
different modes which makes the use of a flux limit survey a rather simplified approach. 
The same applies to possible discoveries of fainter TDEs with longer integration times, by applying the image mosaics technique \citep{krimm13}. These have to be promptly followed-up by XRT for their identification: monitor the soft X-ray emission and then measure the characteristic temporal slope of TDEs.
In this way, a further efficiency accounting for any reason preventing XRT to monitor the event has to be included in our rate calculations (e.g., the stochastic nature of the Swift pointing plan, the target visibility and the mission schedule; Krimm, private communication). Such an efficiency is hard to quantify and any assumption would be arbitrary and would bias our discussion on the comparison between the predicted and observed rates. In addition to that, our soft X-ray modelling assumes a total disruption of the star (see \S2), while the Sw J2058 emission seems to be consistent with a partial disruption. For both of these reasons, we focus on detections triggered onboard, although with due \textit{caveats}. 
In fact, any reliable prediction based on on-board triggers would require complex simulations as 
done by \citep{lien13} for GRB rates. We therefore set to achieve a less ambitious aim at predicting
indicative rates, which should be considered most likely as upper limits.
Specifically, we adopt the BAT daily sky coverage reported in \citep{krimm13} and fix a unique 
``survey'' flux limit to be consistent with the detection of the Sw J1644. We 
detail the procedure in the following.

Sw J1644 was detected with an on-board BAT image trigger \citep{cumming11}. 
In this trigger mode, we assume a flux limit of $2.5\times 10^{-9}$ erg cm$^{-2}$ s$^{-1}$ in the $15-150$ keV 
band, which is consistent with the faint tail of the observed GRB rate \citep{lien13} and the 
detection of Sw J1644 \citep{burrows}. We adopt a daily sky coverage of 85\% \citep{krimm13} and 
apply an efficiency of $\sim 90\%$, for the fraction of the BAT survey time \citep{lien13} 
resulting from trigger deadtimes (e.g. due the passage through the South Atlantic Anomaly).

The detection rate is estimated by performing a large set of MCs as described in
sections \ref{sec:X-ray_modelling} and \ref{sec:MC}.  In eq.\ref{eq:lx}, we use a radiation 
efficiency of $\sim 0.3$ \citep{burrows}. 
For each event, we compare the flux at the trigger time $\tau$ with our flux threshold.
We obtain a TDE rate  of $\approx 10-20$ events yr$^{-1}$ (see Table \ref{tab:rate_radio_X}). 
The rate distribution with redshift extends up to $z_{\rm max}\approx 0.32$\footnote{Here and in 
the following, we define $z_{\rm max}$ as the redshift at which the expected rate is $0.5$ yr$^{-1}$.}
and  peaks at $z\approx 0.2$. The peak value ranges between 
$ 1 - 5$ yr$^{-1}$      (see Table \ref{tab:rate_radio_X}).       
The peak of the corresponding BH mass distribution is
at $10^{6} M_{\rm \odot}$ and contains $ \sim 23\%$ 
of all events. 

Given our predicted mass and $z$ 
distributions for the observed TDEs, 
an event like Sw J1644 has a chance probability which is a factor of $\sim 10$ lower than that
of an event at the peak rate. Therefore, it is not an unlikely event, but a lower redshift object 
would have had a higher probability. At this point, it is unclear to us if this result is more due 
to our simplified treatment of the BAT trigger, to our assumption of a constant jet luminosity 
for a given BH and stellar mass. Both are very likely to have a role. But since we can not trust at this level 
our trigger modelling and the paucity of detected events does not constrain a possible 
luminosity function, we do not attempt here to modify our X-ray model to fit this observed 
distribution. When more events will be identified, our procedure can be refined to account for 
a TDE variety. A comparison with future, easier to 
model, surveys (see Sec. \ref{sec:futureX}) will also help constructing a more robust emission 
model.

Interesting, these rates  are actually up to  two order of magnitude higher than that (0.3 yr$^{-1}$) 
derived from BAT observations,  but a key role is played by the low value of $\Gamma$ considered. We will elaborate on this point in section
 \ref{sec:discussion_current}.

\subsubsection{Radio surveys at 1.4 GHz}

We compare our predictions with constraints on the jetted TDE rate derived from current
 radio surveys \citep{bower}. In the following radio estimates, we will require a 5-$\sigma$ flux limit to claim detection.

We first consider the combined catalogs of VLA First and NVSS at 1.4 GHz.
The combined sky coverage is 0.19 sr with a flux limit of 6 mJy at 1.4 GHz. The analysis of 
these catalogs didn't yield any TDE candidate.

To derive our predictions, we adopt the radio modelling described in section 2.3, and for 
each event (i.e. for each set of $\tau$,
 black hole and stellar mass, and redshift), we calculate the {\it average} flux over a period of one day from the trigger. 
This is compared with a 6 mJy flux threshold. 
 Rescaling our all sky results for the catalogue sky coverage, we obtain an observed rate
 that even in the most favorable case
  ( $\sim 0.3$ yr$^{-1}$ using BM model, eq.\ref{eq:Lvradio}) is consistent with \cite{bower} and 
 \cite{frail} results.   To  strengthen this conclusion, 
  we note that our assumption of a 6 mJy threshold per day combined with a sky
 coverage of 0.19 sr may be considered already rather optimistic, since both values are referred to a 1 yr single epoch.

In the near future, the VLA Stripe 82 survey  may constrain jetted 
TDE models thanks to the improved sensitivity (50 $\mu$Jy  rms) at 1.4 GHz over a FoV of 90 deg$^2$ 
\citep{hodge13}. By assuming a 5-$\sigma$ threshold of 0.25 mJy, our modelling
predicts a number of a few objects to be detected per year.
Significant advances in TDE detections are expected to come from on-going wide radio surveys at both low (see e.g. MWA and LOFAR) and high radio frequencies 
(e.g. ASKAP and VLASS). Since our radio modelling was constrained by observations at higher ($> 1.4$ GHz)
radio frequencies  (as discussed in \S2.3), we focus here on ASKAP and VLASS.  
The Variable and Slow Transient (VAST) project on ASKAP envisages a sky coverage 
of $10^4$ deg$^2$ reaching a sensitivity of $0.5 ~\rm mJy ~\rm rms$ (VAST wide) with a
 daily cadence \citep{murphy}. 
Our predictions for VAST are in the range of a few  up to $\sim 14$ TDE yr$^{-1}$, consistent with expectations from \cite{murphy}.
For comparison, \cite{frail} obtain a value of $\sim 82$ yr$^{-1}$ by considering longer integration time ($\sim$ ten days).
In the case of the  VLA Sky Survey  \citep[][VLASS]{hallinan}, we consider a sky coverage of 
$10^3$ $\rm deg^2$ with $\sim 3$ week cadence at a sensitivity of   $0.1~\rm mJy ~\rm rms$.
 This set up should give a number between 2 and 6 objects to be detected per year.  All sky VLASS  
is also foreseen and will clearly provide a larger number of TDEs, but we focus on the previous strategy 
because the multi-epoch survey could provide alerts  for follow-up at higher energies, with a prompt
 identification of the transient.

\subsection{Future instruments}
Currently, the only two jetted TDE candidates were discovered in X-rays,  where the characteristic $t^{-5/3}$ decay slope has been observed. Therefore, we first discuss 
the discovery potential of 
future X-ray surveys.
We then predict the expected rate of TDEs for the SKA 1.4 GHz wide survey. Finally, 
we derive the properties and rate of TDEs 
that can be detected in radio with SKA and subsequently identified in X-rays.

\subsubsection{Future X-ray surveys}
\label{sec:futureX}

The rate estimates provided in this section are based on a unique observing strategy aimed at detecting and providing a first identification of the transient as a TDE. 
We  assume that a given fraction of the sky is covered in 1 day at a flux threshold defined by the requirement to follow the typical TDE decay over 4 lightcurve bins, each with $\rm S/N \ge 5$.  This is obtained by starting from the 5-$\sigma$  flux limit of each survey, then tracing back  the $t^{-5/3}$ decay in order to obtain the flux over the 4 time bins and then compute the average flux over that period.  This average flux defines the \textit{identification} flux threshold.
We will give values for both $\Gamma=2$ and $\Gamma=20$ and we will justify this choice and elaborate on the comparison in section \ref{sec:discussion}. 

The all sky survey mission eRosita \citep{merloni12} is expected to detect jetted TDEs,
 in its ``hard" X-ray band ($2-10$ keV). 
We apply our methodology to the eRosita survey, properly re-scaling the sky coverage achieved in 
a 6-month scan to 1 day. 
We derive the \textit{identification} flux threshold  for our observing strategy from the $2-10$ keV 5-$\sigma$ sensitivity of  $\sim 10^{-12} ~\rm erg ~\rm cm^{-2} ~\rm s^{-1}$, corresponding to $\sim 250$ s 
exposure \citep{merloni12} as foreseen for each point 
in the sky. 
We calculate the  corresponding un-absorbed flux
 and then we extrapolate it in the energy range $1-10$ keV (used in our X-ray modelling).
We predict a maximum of $\sim15$ TDE per year to be detected up to $z\approx 2.5$, although $z_{\rm max}\approx 0.4$. 
The peak rate is between 0.15 and 0.5 yr$^{-1}$ at $z \approx 0.4$ and beyond $z\approx 2$, the 
rate is $< 5\times 10^{-2}$ yr$^{-1}$. If a larger value of $\Gamma=20$ is considered, the rate decrease by two orders of magnitude
(see Table \ref{tab:rate_radio_X}) with a maximum total rate of $\sim 0.15$  yr$^{-1}$ and peak rate of only $4\times 10^{-3}$ yr$^{-1}$.
We therefore predict both higher ($\Gamma=2$) and lower ($\Gamma=20$) rates than  those previously published by \citet[][1 object to be detected per 6-month 
 long scan]{khabibullin}, but we definitively reach a much lower redshift ($z_{\rm max}=0.4$ vs their $z=4.5$).   The same authors provide 
an upper limit of $\approx 150$  events per scan by considering the number of jetted TDEs to be a 1/5th of their ``soft''
 TDE sample ($\approx 1000$).  This fraction is based on results from the Rosat X-ray survey 
\citep{donley}. We are clearly consistent with their estimate.

The Wide Field Monitor (WFM) aboard of a LOFT-like \citep{feroci} mission will
 also have the capability to trigger jetted TDEs by surveying 1/3rd of the
 sky with a 5-$\sigma$ 1 day sensitivity of $\sim 8-9 \times 10^{-11}  ~\rm erg ~\rm cm^{-2} ~\rm s^{-1}$
 (a few mCrab) in $2-50~\rm keV$ energy band.  We estimate tens of objects per year up 
to $z_{\rm max}\approx 0.6$.
 The peak rate is $\sim 6$ yr$^{-1}$ at $z=0.2$. These numbers imply a total rate of 0.7 yr$^{-1}$  for $\Gamma=20$ with a peak rate of only $6\times 10^{-2}$ yr$^{-1}$.

Finally, we consider Einstein  Probe,
   which is expected to monitor $1/2$ of the entire sky in the energy range $0.5-4$ keV with a 5-$\sigma$  sensitivity 
   of $\sim 10^{-11} {\rm erg} ~{\rm cm^{-2}} ~{\rm s^{-1}}$ in each point (1 ks exposed) of the sky  (EP, W. Yuan private communication).
 In this case, MCs were adapted in order to extend our X-ray modelling to this energy range. This requires to first estimate
   the un-absorbed flux limit by accounting for both the Galactic and the intrinsic absorption
 \citep{burrows} and then
    calculate a proper radiation efficiency by extrapolating from the value inferred in 1-10 keV. We 
estimate a number between $\sim 90-240$  yr$^{-1}$ to be detected below
 $z_{\rm max} \approx 1$ with a peak rate 
    of $\approx 15$ yr$^{-1}$ at $z = 0.3$.  In the case of $\Gamma=20$, a few objects are expected to be detected per year, with a peak rate of $\sim 0.2$ yr$^{-1}$.
    A summary of the actual numbers can be found in Table \ref{tab:rate_radio_X}. 
    
Inspecting the trigger time distributions (see top  panel in Figure \ref{fig:triggers}), we find that up to $z_{max}\sim 1$ objects are detected with almost equal  probability at any delay from the explosion.   

These X-ray survey rates have been obtained under the assumption of a reasonable observing strategy. 
A larger sample extending up to higher redshift can be obtained if longer integration times are considered, 
but these predictions are affected by several parameters like the trade-off between sky 
coverage and sensitivity. In this respect, our approach has to be considered conservative.

\subsubsection{SKA as TDEs hunter}
\label{sec:ska}
Presently, the most ambitious and revolutionary project in radio astronomy is the Square Kilometer
Array \citep[SKA][]{ska2004} planned to operate in 2020. SKA, in survey mode (SKA1-Survey, \cite{Dewdney13}), 
is able achieve a half sky coverage  (20,000 deg$^{2}$) with a 2-day cadence at a $5-\sigma$ flux limit of  $90 ~\mu \rm Jy$ \citep{donnarumma14, feretti2014}. 
These unprecedented sky coverage and sensitivity make SKA an optimal radio transient hunter.

Differently from X-ray searches, in radio, we cannot have a first identification based on the lightcurve, since the 1.4 GHz radio emission of a TDE is not particularly different  from those of other radio transients (e.g. GRB, blazars). Therefore, we consider a different strategy. In our MC simulations, we  directly assume the SKA $5-\sigma$ flux limit in order to claim the detection of a transient event. 
The {\it identification} strategy will fully rely on the multi-frequency follow-up of the trigger event as it will be discussed at the end of this section.

\begin{figure}[]
\epsscale{1.7}
\plottwo{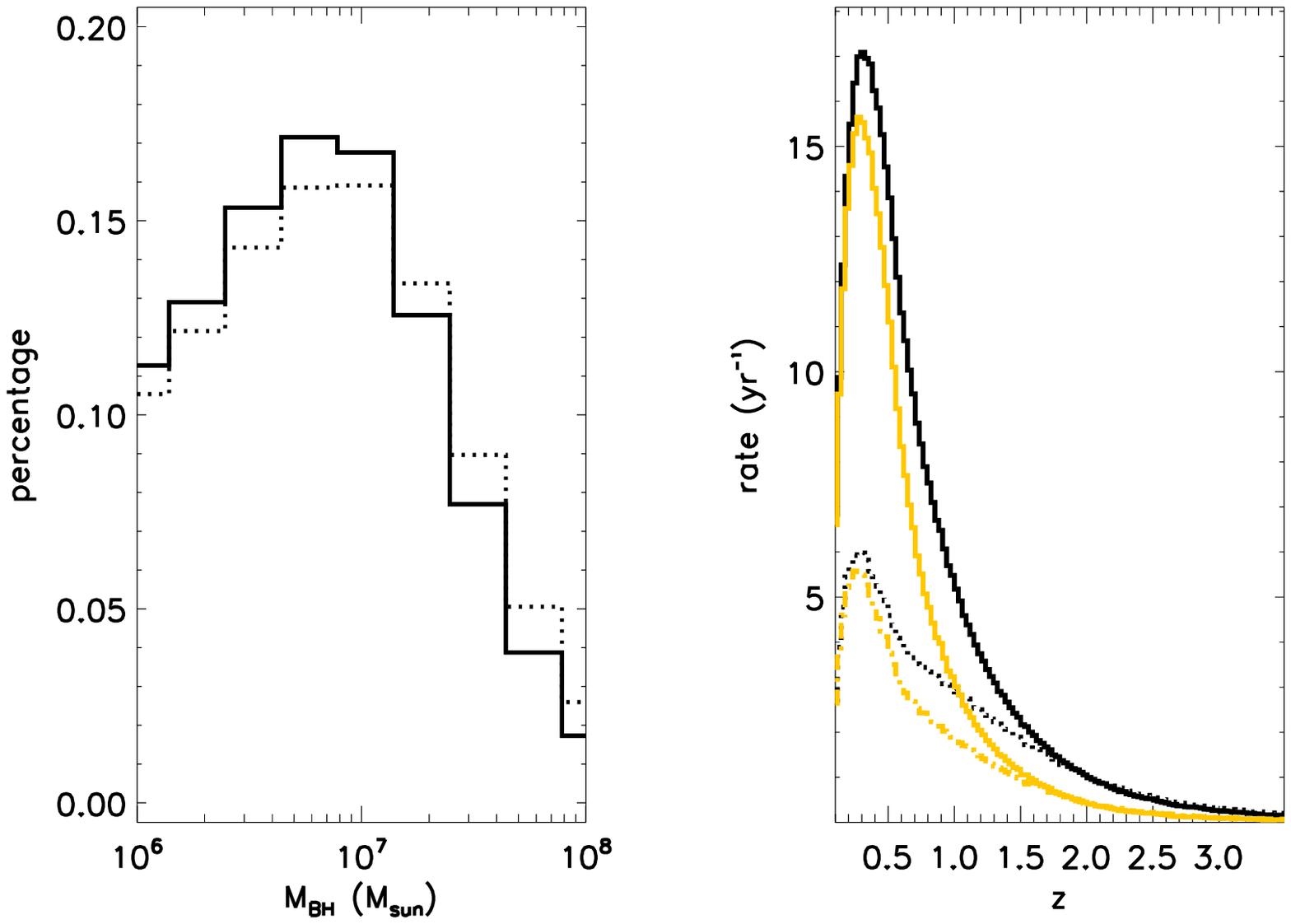}{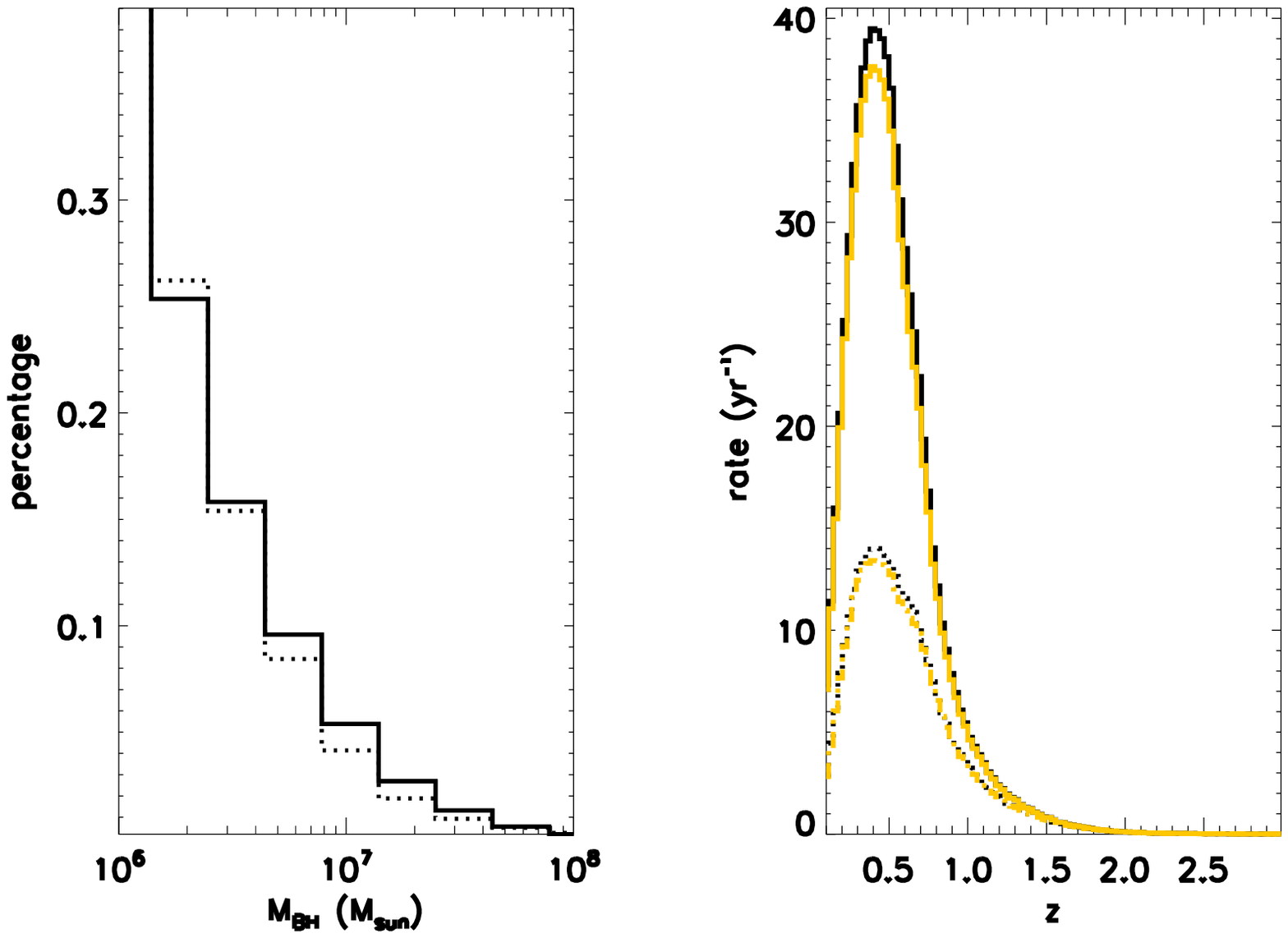}
\caption{Rate of events predicted for SKA in wide survey mode at 1.4 GHz as a function of redshift 
(right panels) and their distribution as a function of BH mass (left panels), for two different black hole distribution functions (black solid line:   G model
black dotted line:  Gz) model. Rates  associated to
events with BH masses lower than $10^{7}$ M$_{\odot}$ are also shown (yellow lines). {\it Upper panels:} BM model for 
the jet evolution;  {\it lower panels},  MDL model. }
\label{fig:rate_radio}
\end{figure}

We calculate the predicted average flux over 2 days from the trigger and then
we compare it with the SKA flux limit.
The results are shown in Figure \ref{fig:rate_radio}. The upper panels are derived using the BM model
(eq.\ref{eq:Lvradio}) for the radio lightcurve modelling, while the lower panels use the MDL model
(eq.\ref{eq:Lvradio_Mbh}). There, we show the distribution of the TDE rate as a function of $z$
(right panels) and their BH mass distribution (left panels) for the two BH mass functions described
in \S 4 (black lines). The yellow lines show the subclass of events with BH masses lower than $10^{7}$
M$_{\odot}$. Both radio models produce redshift distributions peaking around $z\approx 0.4$, regardless of the BH mass function. The peak rates are  roughly between $6$
and 40 yr$^{-1}$ (see also Table \ref{tab:rate_radio_X}). Events with BH mass lower than $10^{7}$ $M_{\sun}$ dominate the distributions
at all redshifts in the MDL model, while this only happens at $z < 0.4$ in the BM model. One
marked difference between the two radio lightcurve modellings is the BH mass distribution of the
detected events: while BH with masses between $10^{7}-10^{8} M_{\sun}$ are equally probable in the
BM model (because the flux is BH mass independent, eq.\ref{eq:Lvradio}), BHs with mass $< 3 \times 10^{6}$ completely dominate the observed
sample in the MDL model. As a consequence, BM model distribution extends to higher redshifts ($z_{\rm max} \approx 2.5$ vs $z_{\rm max} \approx 1.7$), 
because  BH masses larger than $10^7$ M$_{\rm \odot}$ interact with higher mass stars (higher $m_{\rm *, min}$) and produce intrinsically brighter flares. This will allow us to  study TDEs close to the
peak of cosmic star formation.

In Table \ref{tab:rate_radio_X}, we also report the total rates obtained by integrating
these distributions in $z$ and M$_{\rm BH}$.
We obtain yearly rates of  the order of a few to several hundreds.   
These results are {\it not}  consistent with those that can be derived by using eq.4 in \citep{vanvelzen}: inserting our SKA survey parameters, we obtain thousands of events per year.
This discrepancy is due to our inclusion of the stellar mass dependence, that modulates the TDE luminosity for a given BH mass: the lower $m_*$, the dimmer the event.
 In the assumption of a Kroupa IMF, the bulk of the events are caused by the disruption of stars with $m_{*} < 1$, increasing the number of flares that are too dim to be detected.

With hundreds of events per year, SKA could be able to detect  more TDEs than any currently planned X-ray survey. 
On the other hand, while X-ray surveys can catch the
events soon after explosion (see EP performance in Figure \ref{fig:triggers} upper panel, for an example), SKA would not be able to cover the first week activity at any redshift and 
only at $(z < 0.8)$ SKA will probe the first month (Figure \ref{fig:triggers}, bottom panel). This result is independent on the assumed radio modelling.
The explanation is simple:  the observed radio
flux at 1.4 GHz is initially increasing, contrary to that in the X-ray band. In this regard, detections in these two bands are complementary.
However, a word of caution  here is due. As mentioned before, below 10 days, we have virtually any detection in radio at any $z$. This early period  coincides with the
 rise of the radio light curve. Although we expect this gap in detection, the exact epoch at which 
 it occurs depends on the detailed behavior of the light curve during this undetected rise. Our extrapolation 
 at earlier times is quite steep and we consider 10 days as an upper limit for the initial gap in detection.

So far we focused on {\it detection} of TDEs with SKA, that, depending on the observing strategy, will only be a fraction of a noticeable sample
 of \textit{slow} radio transients. As mentioned earlier, we cannot use  radio properties or variability alone to distinguish a
TDE candidate from neither a slowly variable AGN or a GRB.  A possibility for {\it identification} that we explore below is through quick follow-ups at higher energies, particularly in X-rays.
A first pre-screening of the radio candidates could be done by 
 cross-correlating the radio transient positions with deep AGN catalogues, expected to be provided in the near future by optical surveys (e.g. LSST) 
or the SKA precursors (e.g. ASKAP). However, we expect a larger degree of contamination of the TDE sample to come
from transient sources such as GRBs. Since, unlike GRBs, most of TDEs should have a nuclear origin,
 it is mandatory to quickly identify the host galaxy. An accurate localization of the radio transient in the core of galactic nuclei, helping to assess the nuclear origin, will therefore play a major role in the screening of the radio transient sample. This means  that first the host galaxy has to be  found by a rapid optical follow-up and after the brighter transients could be localized by SKA with a precision of $\sim 100$ milliarcsecond\footnote{this can be achieved thanks to the resolution of about 2 arcsec of SKA1-SUR and 0.6 arcsec or better of SKA1-MID \citep{Dewdney13}} (mas) essential to separate nuclear transients from other phenomena (e.g., GRB). 
 
For details see \cite{donnarumma14}.

\subsubsection{Combining Radio and X-rays in the SKA era}

X-ray follow-up will have a major role in the identification of the TDE candidate detected by SKA because of the possibility to detect the characteristic $t^{-5/3}$ decay.
A possible X-ray follow-up strategy aimed at \textit{identifying} and  then \textit{characterizing} the event consists in a fast repointing of the transient detected by SKA.
We consider a 1-day delay in the X-ray repointing and require a set of X-ray observations spread over a few days in order to follow the characteristic temporal decay of the TDE. We foresee an observing strategy which is similar to the one adopted in the case of future X-ray surveys (see section \ref{sec:futureX}): four observations spread over \textit{4 days}, 
with $S/N$ ratio $\geq 10$ in each.  A high $S/N$ is required in order to characterize both the temporal and spectral behavior of the source.     

For each event in the MCs, we calculate the average X-ray flux over the 4 days after the repointing
and compare it with the \textit{identification} flux threshold derived as explained in section \ref{sec:futureX}, with the only difference of a $S/N \ge 10$ requested in each observation.
Practically, $\tau$ in eq.\ref{eq:lx} has to be the radio trigger time-lag, plus an extra
delay of one day for repointing, and $0 \leq \Delta t \leq 3$. In this way, we derive the properties of samples of TDEs which
are first  detected in radio and promptly followed-up in X-rays.
 
In Fig. \ref{fig:rate_x_flux_limit}, we show the fraction of SKA candidates that can be identified as a function
of the X-ray (1-10 keV) unabsorbed flux limit. A rapid X-ray follow-up will be able to detect a complete radio-selected sample provided that the
instrument sensitivity is close to $F_{\rm lim} \lesssim 10^{-11} ~\rm erg~cm^{-2}~s^{-1}$ in the energy. 
In fact, a moderate sensitivity  $\sim 10^{-11} - 10^{-10} ~\rm erg~cm^{-2}~s^{-1}$ is already enough to detect equal 
or a larger number of events than with X-ray {\it wide} sky instruments alone. It is therefore clear that a radio trigger 
is a more efficient way to build up a large X-ray sample of TDEs.
Rates reported in that Figure assume a fast (1 day) X-ray repointing and $F_{\rm lim}$ reached with an integration of $\sim 4$ days.
 Rates could be substantially different if longer integrations are needed to reach the same $F_{\rm lim}$ or in the case of longer repointing time. 
This is a natural consequence of the decreasing trend of the X-ray light curve.

When considering an actual follow-up strategy,  the values reported in
Figure \ref{fig:rate_x_flux_limit} should be scaled by the fraction of sky accessible to the X-ray
instrument considered. In general, the X-ray follow-up will provide us with a sub-sample of radio
triggered TDEs, defined by the target accessibility, the repointing chance of the X-ray satellite
and the sensitivity of the instruments. Since TDEs also emit in hard X-rays, a trade-off between sensitivity, sky
coverage and a broad energy range is foreseen. In particular, the broader is the energy range the
better the characterization of the non-thermal process and of the jet energy budget.

Future X-ray experiments like Athena \citep{nandra2013} and a LOFT-like mission  \citep{feroci}
 could offer a unique chance to follow-up and characterize SKA triggered TDEs. Moreover,
 if Swift were still operating in the 2020s, XRT will have a great potential in following-up the
 radio candidates.

Athena sensitivity lies in the saturation branch of Figure \ref{fig:rate_x_flux_limit}, which implies
that the observed rate of X-ray jetted TDEs will be crucially linked to its follow-up efficiency.
This is mainly influenced by the Athena sky accessibility which is of the order of $\sim 50\%$ 
\href{http://www.the-athena-x-ray-observatory.eu/}{(Athena mission proposal)},
resulting in a rate of TDEs of a few hundreds, with detections up to  
$z_{\rm max} \approx 2$ (see Table \ref{tab:rate_radio_X}).

The LAD (Large Area Detector) on board of LOFT (2-50 keV) is a collimated instrument with 1 degree
field of view, and a background limited sensitivity of $\geq 10^{-12} ~\rm erg ~\rm cm^{-2} ~\rm s^{-1}$ 
in the 2-10 keV band, for a 100 ks exposure. The LOFT pointing visibility will assure a sky accessibility for these targets of $\sim 75\%$,
\href{http://sci.esa.int/loft/53447-loft-yellow-book/#}{(LOFT Yellow Book)}.
The requirements of our strategy define a $F_{\rm lim} \sim 10^{-11}
~\rm erg ~\rm cm^{-2} ~\rm s^{-1}$ in the 2-10 keV band, which was then translated in the
corresponding un-absorbed value in the 1-10 keV band (the energy range adopted in our modelling).
Again, we assume a 1-day repointing delay.
Figure \ref{fig:rate_X}, shows the expected rate of jetted TDEs for a LOFT-like mission as a function
of redshift (right panel) and their mass distribution (left panel). The rate distributions
are calculated under the BM model (top panels) and MDL model (bottom panels) for the radio 
modelling.
 In both cases, we found that the redshift distribution extends above $z =1$ ($z_{\rm max} \approx 1.2 - 1.7$) 
 (see Table \ref{tab:rate_radio_X}), with most of the TDEs expected around $z\backsimeq 0.4$
(right panels). The peak rates are roughly between  $\approx 4$ and 20 events per year. In the MDL
model,  because of the mass dependence of both the radio and X-ray luminosities, $\sim 25\%$  of all
 events have BHs with masses $\approx 10^{6} ~\rm M_{\odot}$, and events with BH masses $ < 10^{7}
~\rm M_{\odot}$ dominate the redshift distribution at all epochs.
In the BM model, instead, the TDE rate peaks at $\approx 10^{7} ~\rm M_{\odot}$ (see left panel in Figure \ref{fig:rate_X}),
with lighter BHs dominating at $z \le z_{peak}$ (yellow lines in  Figure \ref{fig:rate_X}, right panel). 
The behavior at higher $z$ fully reflects the one observed in the BM radio rates (see  fig. \ref{fig:rate_radio}). 
In total, a LOFT-like mission should be able to detect a sub-sample of radio TDEs between
$\approx 130$ and $\simeq 350$ yr$^{-1}$.  Instead, very few objects per year are predicted if $\Gamma=20$.
For these events, the
mission broad energy  band ($2-50 ~\rm keV$) should enable us to put tighter constraints on
the energy budget of the X-ray component, than possible with Athena instrument.

The price to pay for detecting more X-ray TDEs with a follow-up strategy is illustrated in Figure
\ref{fig:triggers} upper panel, where
we compare the trigger distribution for the EP (black lines) and the LOFT radio triggered (blue
lines) samples. Most of LOFT events are observed after 10 days from the beginning 
of the emission\footnote{See discussion in Sec. \ref{sec:ska}}. In
particular, high redshift $z \approx 1$events are all a couple of months old. Direct discovery of
TDEs in X-rays is thus important for catching the event in its very early dynamical stages, when the
jet has just formed and the disc may still be in the (largely unconstrained) super-Eddington regime.


\begin{table}
\begin{center} 
\begin{tabular}{c|c|c|c|c|c|c}

& $R^{1}$  & $R^{2}$ & $z_{\rm peak}$ & $R_{\rm peak}^{1}$ & $R_{\rm peak}^{2}$& z$_{\rm max}$ \\
&    yr$^{-1}$ & yr$^{-1}$ & &yr$^{-1}$ &yr$^{-1}$ &\\
  \hline
  \hline
  Radio selected sample & & & & &\\
\hline
  SKA  {\it BM}    &226   & 468 &$ 0.3$ & 6 & 17 & $ 2.5$\\
  SKA  {\it MDL}    & 327 & 770  & $  0.4 $ & 14 & 40 & 1.7\\
  LOFT-like  {\it BM}     & 128 (2.5) & 305 (6.5) &$ 0.3$ & 4.5 (0.05) & 13 (0.1) & 1.7 \\
  LOFT-like  {\it MDL} & 135  (1.3)& 352 (3.5) & 0.4 & 8 (0.08)& 22 (0.2)& 1.2\\
  Athena {\it BM}     &113 (1)& 234 (2.3)& 0.3& 3 (0.03) &8.5  (0.09)& $ 2$\\
  Athena {\it MDL} &  163 (1.6)& 385 (4) & 0.4 & 7 (0.07) & 20 (0.2)&  1.4\\
  \hline
  \hline	
  X-ray surveys & & & & & \\
  \hline
 BAT$^{3}$    & 9.5 (0.095)  & 26.5 (0.26) &$ 0.1-0.2$ & 1.7 (0.02)& 4.6 (0.05)& 0.32\\
eRosita &  8 (0.08) & 15 (0.15) & $ 0.4$ & 0.15 (0.001) & 0.5 (0.005) & 0.4\\
Einstein Probe      &89 (0.9)&  242 (2.4) & 0.3 &5.5 (0.05)  &15 (0.2)  &1  \\
LOFT-like WFM    & 24.5  (0.2) & 67 (0.7) &0.2 & 2.3 (0.02) & 6 (0.06) & 0.6 \\
\hline
  \end{tabular}
\end{center}
\caption{Future Radio and X-ray surveys predictions: 1st and 2nd columns are total 
yearly rate
(the subscripts $^{1}$ and $^{2}$ are for the $Gz$ and $G$ MFs, respectively), 3rd column redshift 
at the peak rate, 4th and 5th columns maximum  peak rate and 6th column maximum 
redshift, defined as the z where the rate is 0.5. BAT$^{3}$: calculation for an on-board image trigger. X-ray and Radio expected rates are derived for $\Gamma=2$. X-ray rates are also reported for $\Gamma=20$ in parenthesis.}
  \label{tab:rate_radio_X}
\end{table}

\section{Discussion}
\label{sec:discussion}
The Swift/BAT discovery of  Sw J1644 opened a window on a new class of X-ray and 
radio transients, which are optimal targets for future radio and X-ray surveys/instruments.
The study of these objects allows us to investigate the formation of transient jets in extra-galactic sources.
 Moreover, there is the potential to discover quiescent SMBHs in distant galaxies and constrain the
 SMBH mass function.
In this section, we qualitatively discuss our results and what we may learn from them. 
Any quantitative parameter investigation (for instance with a Fisher Matrix technique) is beyond the
scope of this present paper, and will be presented in a follow-up work.

\subsection{Jet efficiencies and bulk Lorentz factor}
\label{sec:discussion_current}

 So far, only two jetted TDEs have been detected, while the thermal candidates, 
 related to the presence of an accretion disk, have been more numerous. The question 
then arises whether this is due to observational biases, highly collimated jets or to an intrinsic
low efficiency of transient accretion disks to produce (luminous) jets.

To try and address this question, we could compare our predictions to the Swift/BAT observed rate ($\approx
0.3$ yr$^{-1}$): our lower limit ($\approx 9$ yr$^{-1}$) is a factor of 30 higher. 
It is tempting --- and indeed it has been 
done in the literature --- to reconcile this discrepancy by invoking a jet production efficiency of 
a few percent, since our calculations assume that each TDE is accompanied by a jet.\footnote{In our 
simplified description here, there are only two kinds of possible
events: Sw J1644 with its own jet luminosity (i.e. a given jet energy efficiency $\epsilon_{\rm j}$)
and events with no jet (i.e. $\epsilon_{\rm j}$ \textbf{very} small). In reality, there must be an
intrinsic distribution of $\epsilon_{\rm j}$, with a tail of low energy events that cannot be
detected or failed to be launched at relativistic speeds.} 

However, there are several reasons why this inference should not be drawn.
First, as discussed 
in Section \ref{sec:hardX}, it is absolutely non-trivial to describe  the characteristics (e.g. 
flux limit and sky coverage) of an effective Swift/BAT survey. We believe that our assumptions for the trigger,
together with a  100\% identification efficiency gives rates that are indicative of an 
upper limit. Second, BAT rate predictions, unlike those of other X-ray instruments consider here, strongly depend on the modelling 
of the early stage variability of the X-ray lightcurve (see \S 2).  The 
onboard threshold we use is very close to the flux of the upper envelope of the lightcurve. We are therefore implicitly assuming that we 
can always trigger an event, by catching it at its maximum.
However, since the flux varies by two orders of magnitude, our choice implies again an upper limit estimate of BAT rates.
Finally, even if we trust our modelling of the BAT trigger and initial X-ray variability, uncertainties in the 
value of $\Gamma$  can account for the discrepancy. So far, we have considered a bulk Lorentz factor of 2, 
since the radio measurements strongly support such a low ($\Gamma \sim 2-5$) value.
However, hard X-ray observations are consistent with larger Lorentz factors 
\citep[$\Gamma \le 20$][]{burrows}, which will bring down our rates to the observed value (see Tab.1).
The consequence would be that the simultaneous hard X-ray and the radio emissions need to come
from different regions --- as already claimed \citep[e.g.][]{zauderer}. The picture may be that while 
the radio emission is produced from further out, after the jet has substantially decelerated, 
X-rays probes regions much closer to the central engine \citep{bloom}.
If that was true, X-ray detections and follow-ups would be further suppressed 
with respect to the expected SKA performance.

Unlike the previous comparison with BAT results, our predictions of the radio rates are 
consistent with the upper limits derived using with the NVSS + FIRST catalog \citep{bower}, for any $\Gamma \ge 2$.
As a consequence, this comparison cannot provide us with further constraints on either $\Gamma$ or the jet efficiency.
In the next future, surveys such as VLA Stripe 82, ASKAP and VLASS
will give tighter constraints on jetted TDEs thanks  to the improved sensitivity 
\citep[50 $\mu$Jy rms,][]{hodge13} of the former and the wide field of view of the latter 
two surveys. In this case, our radio modelling predicts a
number of a few objects $\rm yr^{-1}$ (a few tens yr$^{-1}$) to be detected by assuming $\Gamma = 2$. Comparing predictions with
(positive) observations will thus constrain possible combinations of $\Gamma$ and jet production
efficiency.

As already discussed, the optical transient surveys
Pan-STARRS and LSST are expected to make significant advances in the study of TDEs.  LSST will be a
real breakthrough in this respect, surveying $2 \times 10^4$ square degrees of the southern sky. 
{\it Thousands}
of objects are expected to be discovered at $z < 1$ by catching their thermal light from the
accretion disc or from the non-relativistic wind in the Super-Eddington phase,
surveying the same fraction of the sky every 3 days \citep{strubbe2009,vanvelzen11a}.
However, optical extinction in galactic nuclei still introduce an observational bias in the TDEs
discovery although less significant with respect to that occurring in the UV band. As suggested by
\citep{strubbe2009}, infrared surveys will provide a complementary approach being the lower
frequency energy range less affected by any source of obscuration.

Contrary to radio and X-ray emissions, the optical and infrared light are {\em not} expected to be
relativistically beamed nor to be connected with jet emission. These features imply that a
comparison
between optical, X-ray and radio selected samples can help constraining both the TDE efficiency to
produce jets and the relativistic
Lorentz factor. This latter, when an X-ray sample is available, will
help assessing the jet energy efficiency $\epsilon_{\rm j}$.

\subsection{Supermassive BH masses}

To understand supermassive BH cosmic growth and their connection with the host galaxy, it is
necessary to have a good understanding of which mass can be found in which galaxy and, more
broadly, of the SMBH mass function as a function of redshift.

The detection and light modelling of a TDE event is a unique way to constrain the mass of
an otherwise quiescent BH, that is too distant to be detected by stellar dynamics. An attractive
feature is that TDEs may occur in any type of galaxy, allowing for the detection of a broader range of
SMBH hosts. For the lightcurve modelling, a multi-wavelength approach can yield tighter constraints
on the mass, since other parameters such as the jet energy, Lorentz factor and the stellar mass need
to be simultaneously determined.

A perhaps more direct measurement of the BH mass can come from very fast
X-ray variability, as the quasi periodic oscillation (QPO) observed in Sw J1644 \citep{reis}.
The prospect for detection of QPOs in such events is quite favorable for both Athena and a
LOFT-like mission. If QPOs in TDEs were associated with the Keplerian
frequency at the innermost stable orbit  \citep[as discussed in][]{reis}, the highest rest frame
frequency should be of the order of $200 s$ for a BH mass of $10^6 ~\rm M_{\odot}$. This QPO 
frequency is easily within reach of both $LOFT$-like and Athena instruments  
\citep{feretti2014,nandra2013}. Longer oscillations are expected for more massive BHs
($\propto\sqrt{10^6/\rm M_{\rm BH}}$), whose detectability could be more complicated due 
to satellite orbit constraints (e.g. Earth occultation, South Atlantic Anomaly).  However,
 providing that the QPO is persistent over a long period and the source is bright enough to 
remain above threshold for several cycles, a direct measure of such a QPO is also possible.

An other method to constrain the mass function may be to compare our rate distributions with future SKA
triggered observations. As shown by Figure \ref{fig:rate_BH} upper panel, there are still
uncertainties in the BH mass function, which in turn affect our rate predictions (see Figure \ref{fig:rate_BH} lower panel, Figure \ref{fig:rate_radio} and Figure \ref{fig:rate_X}).

\section{Conclusions}
We have investigated the best strategies to increase the sample of the new class of TDEs, which was
recently  discovered by BAT. These events emitted non-thermal emission in X-ray and radio
bands, probing a relativistic jet.
Given the lack of statistics and of a solid theoretical framework for their non-thermal emission, 
we adopted a rather phenomenological approach to model their lightcurve. We fit the behavior of 
the best studied candidate, Sw J1644, in both radio  (1.4 GHz) and X-rays (1-10 keV), and 
we used the classical theory of TDEs to rescale the emission for different black hole and
 star masses. In the radio band, we also considered, in alternative, the blast wave model,
 usually adopted for GRBs.
We then used a Monte Carlo code to compute their expected rate as a function of redshift and black
 hole mass. We considered both current and future radio and X-ray surveys/instruments. 
 Since the characteristic temporal decay of a TDE event can be observed in X-ray, an identification is claimed only when the X-ray emission  can be sampled 
 in at least 4 lightcurve bins with high signal to noise ratio, $S/N \ge 5$. When the TDE is {\it detected} in radio, we investigated a follow-up strategy for {\it identification}
 which required X-ray detectors to sample the lightcurve with the almost the same requirements as above (but with a $S/N \ge 10$ ).  
 To concretely explore future possibilities, we investigated in particular the expected performance of eRosita, Einstein Probe, Athena, 
 a LOFT-like mission and SKA operating in survey mode (SKA1-SUR).  

\begin{figure}
\begin{center}
\epsscale{1.7}
\plottwo{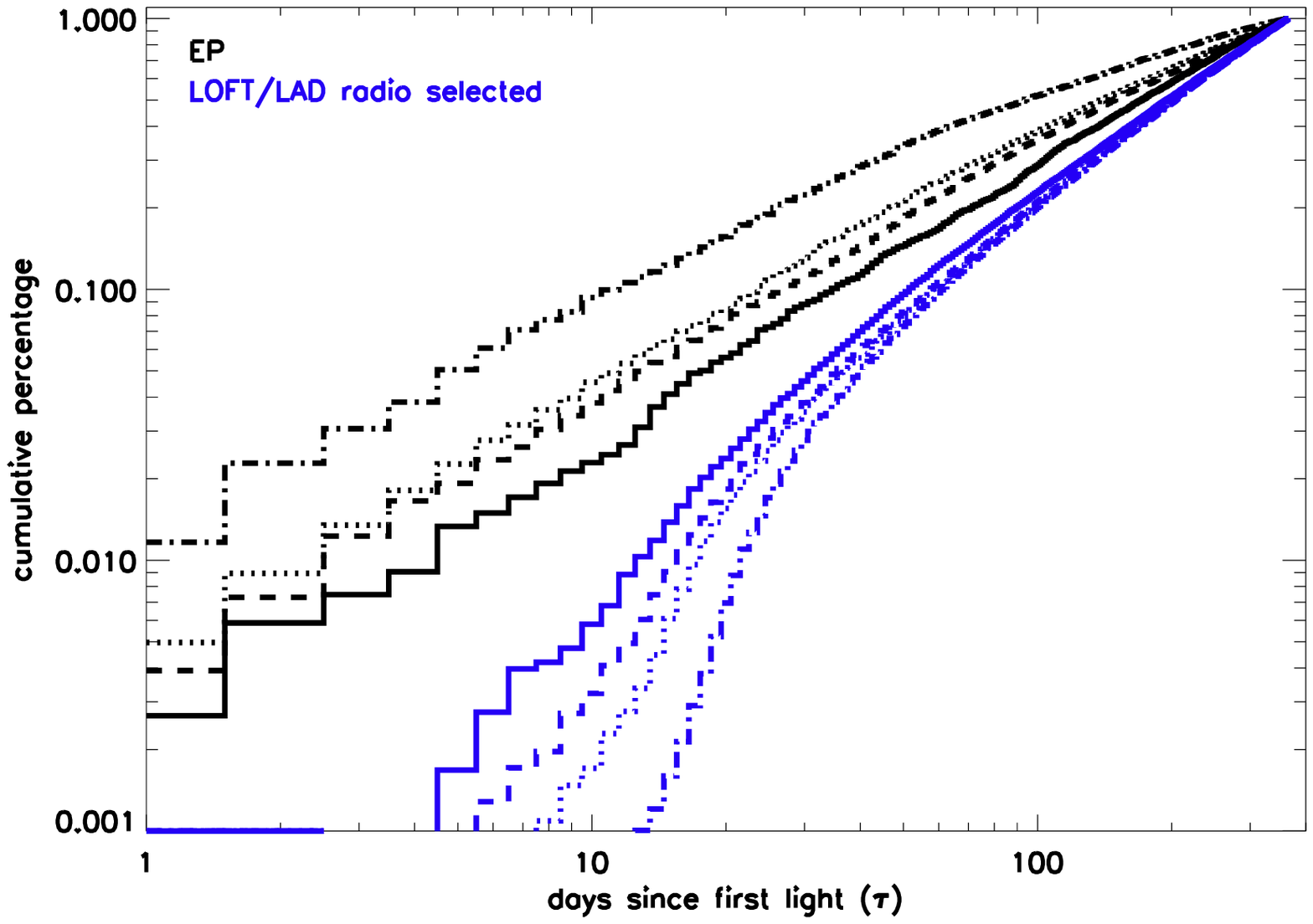}{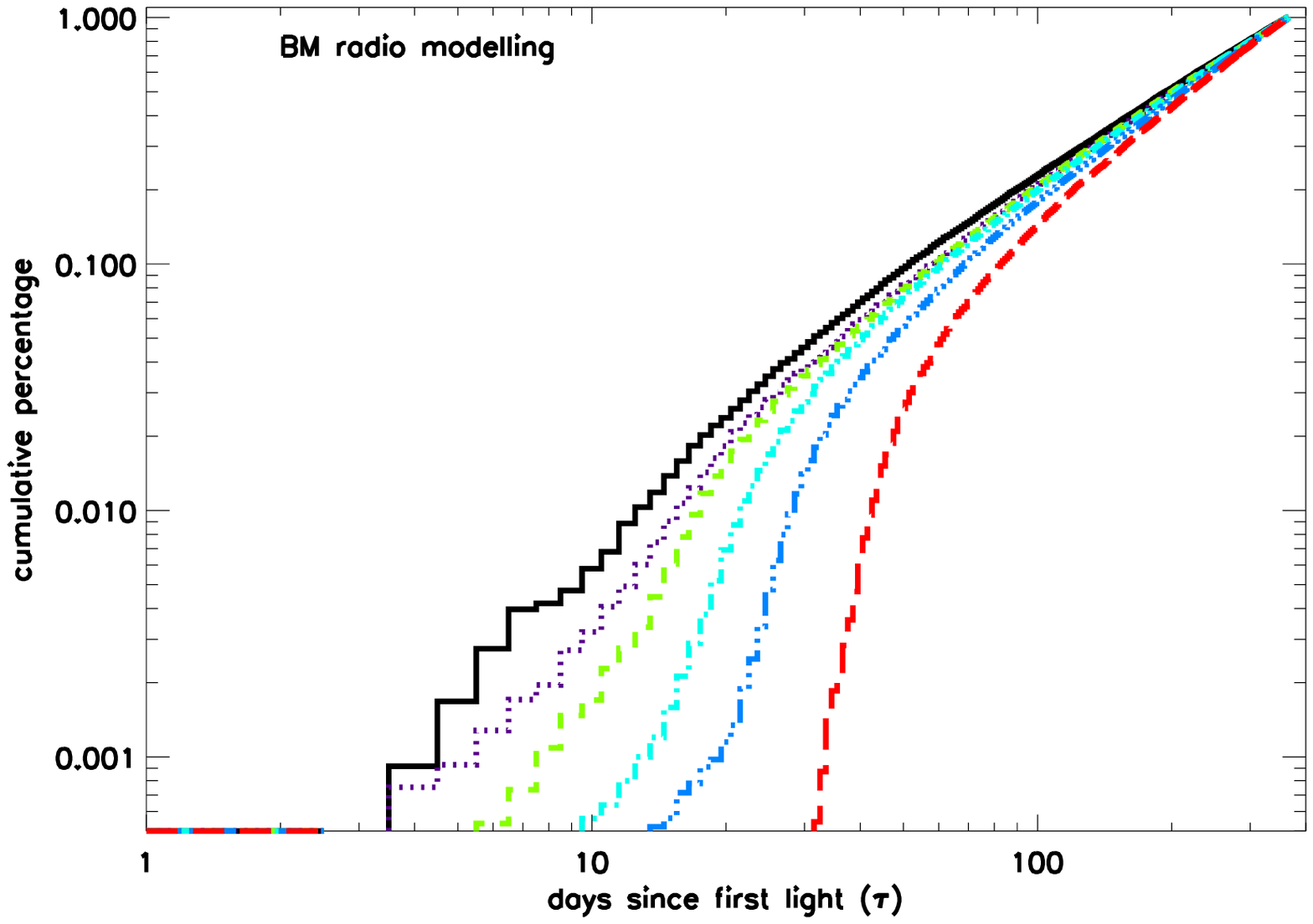}
\caption{Cumulative distributions of delays  in detecting the TDE from the explosion time, for different redshifts. {\it Top panel:}  EP  (black lines) and LAD follow-ups of radio triggered TDEs (blue lines). The different line styles are for $z=0.1,0.2, 0.37, 0.8$ from right to left for EP and viceversa for LAD. {\it Bottom panel:}  the same as above but for SKA BM model and $z=0.1, 0.2, 0.37,  0.8, 1.5, 3$ from left to right). }
\label{fig:triggers}
\end{center}
\end{figure}

\begin{figure}
\includegraphics[width=\columnwidth]{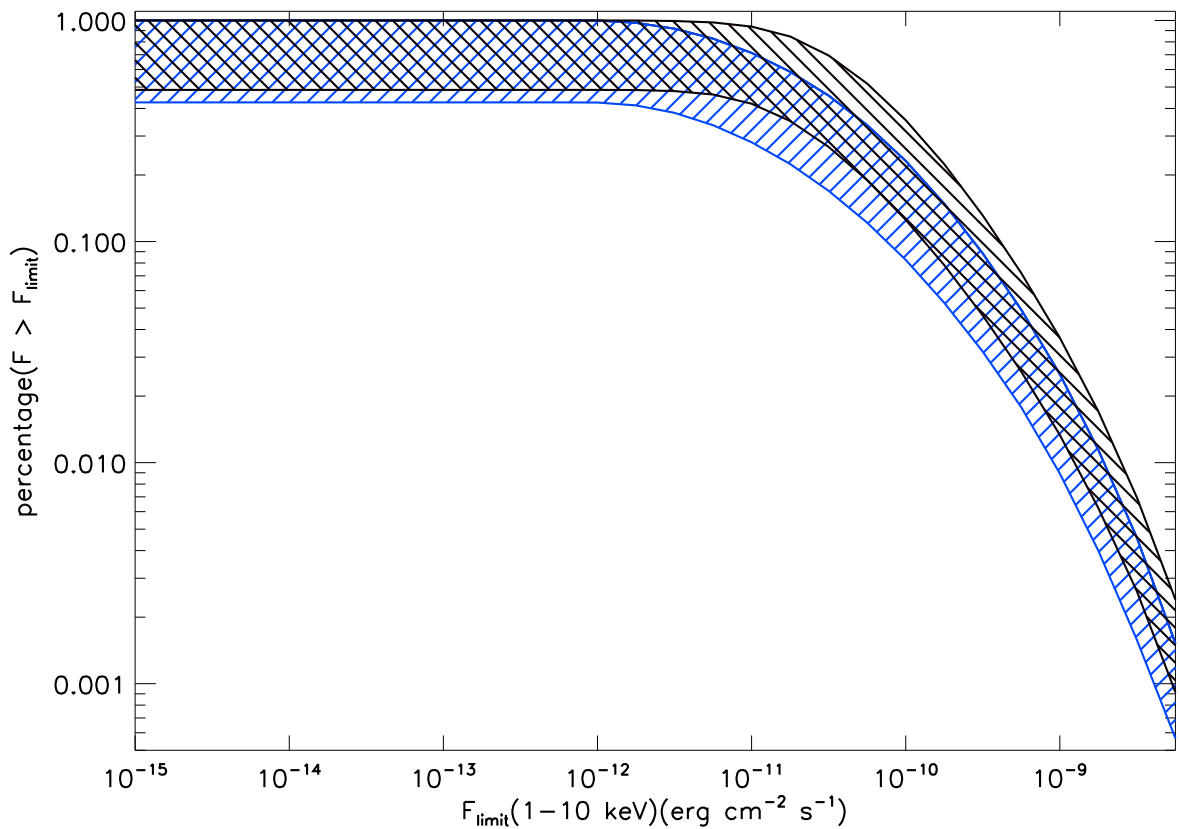}
\caption{Fraction of X-ray TDEs that can be identified, following up a SKA trigger, as a function of flux limit (unabsorbed flux).
The blue and black shaded area are obtained with eq.\ref{eq:Lvradio} and eq.\ref{eq:Lvradio_Mbh} 
radio modellings, respectively. The shaded areas reflect uncertainties in BH mass functions. The figure shows that a X-ray instrument with a
 flux limit of $\lsim 10^{-11}$ in cgs units, can in principle identify any radio detected TDE. See text for details. }
\label{fig:rate_x_flux_limit}
\end{figure}

\begin{figure}
\epsscale{1.7}
\plottwo{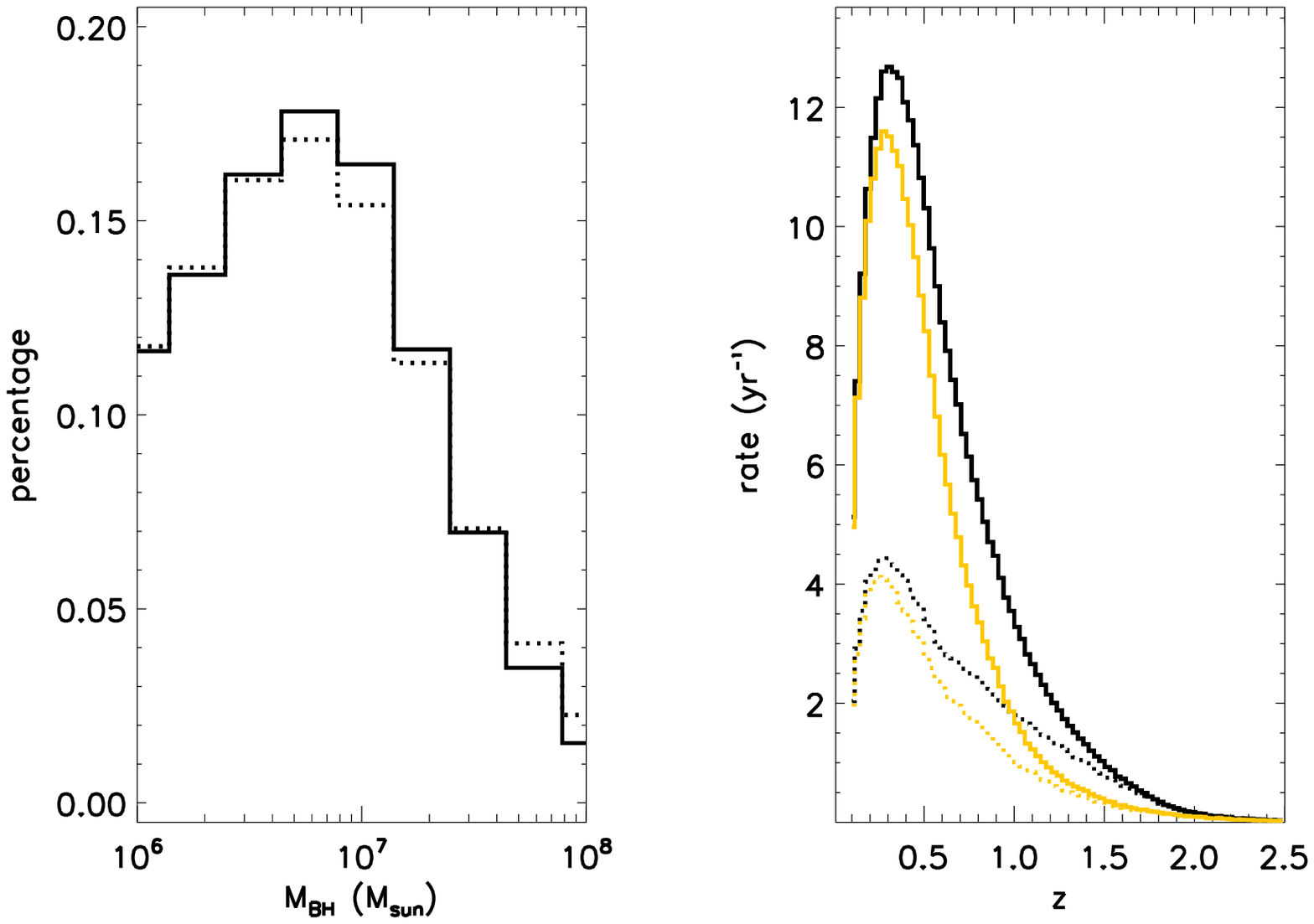}{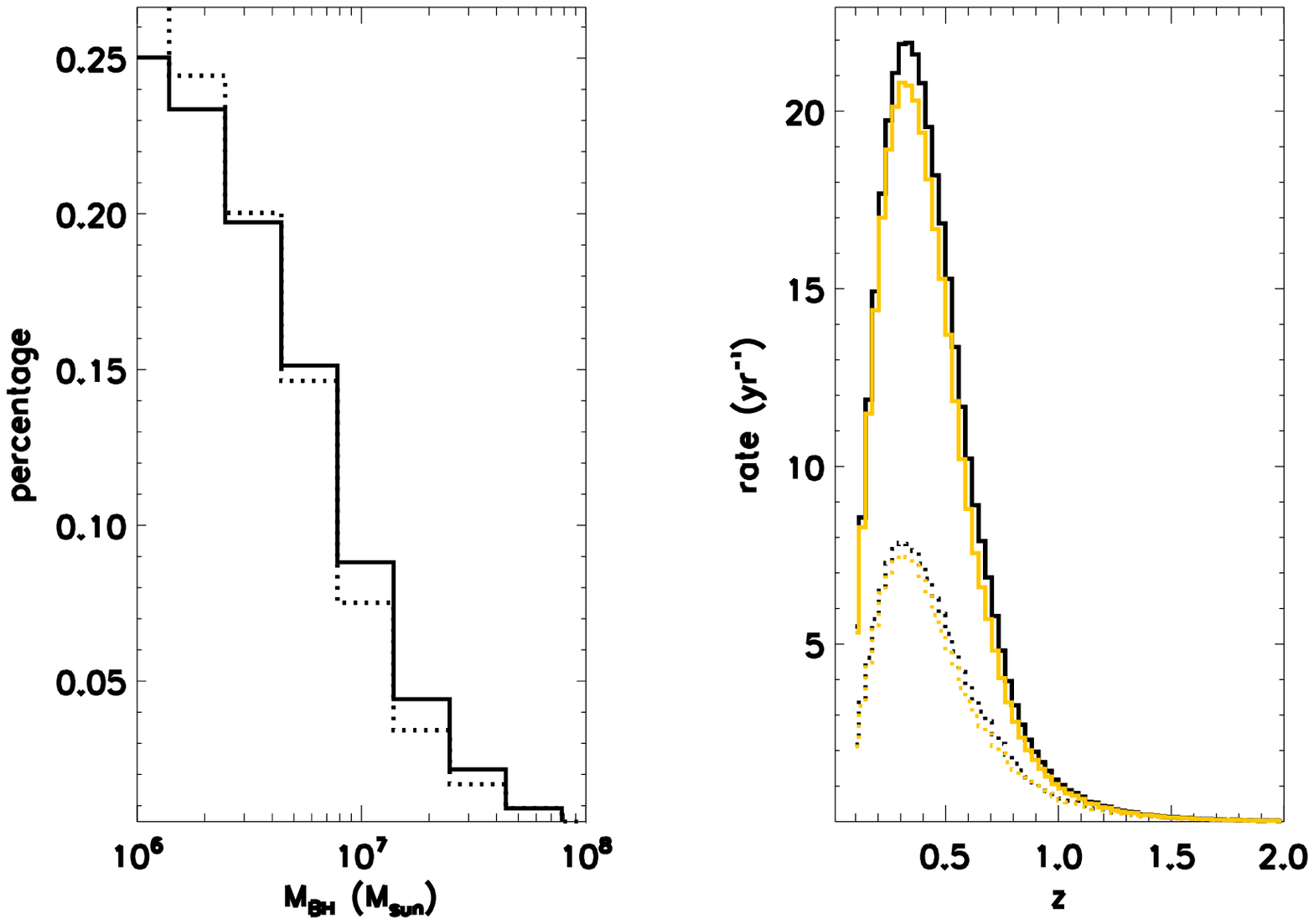}
\caption{As Figure \ref{fig:rate_radio}, but for radio triggered events repointed by a LOFT-like mission.}
\label{fig:rate_X}
\end{figure}

Our major findings can be summarized as follows:
\begin{itemize}

  \item results from current instruments (such as BAT and NVSS + FIRST catalogues) do not provide
constraints on jet parameters or the jet production efficiency;

\item However, to reconcile BAT predictions with observations a $\Gamma \approx 20$ may be adopted, 
consistently with hard X-ray observations \citep{burrows}. If this were true, X-ray and radio emissions should come from two different regions, as already suggested on different bases
\citep{zauderer}. The predicted X-ray rates would also be suppressed by $(2/20)^{2}$ with respect to those in the radio band.

  \item In the near future, VLA Stripe 82 survey, VLASS and ASKAP-VAST may provide from a few to ten events  yr$^{-1}$, 
  putting some constraints on possible combinations of bulk Lorentz $\Gamma$ and jet production 
efficiency;

  \item Hundreds ($\Gamma = 2$) of Sw J1644-like objects per yr are
expected to be within reach of SKA1-SUR  at 1.4 GHz.  They can probe the distant Universe up to $z\sim 2.5$. These results differ from previous,
 more optimistic, predictions of thousands yr$^{-1}$ \citep[for $\Gamma = 2$][]{vanvelzen11} 

  \item Future X-ray surveys will provide a more modest sample,
between several (eRosita) to a maximum of $\approx 240$ (EP) jetted events per year. With a highly collimated jet, with $\Gamma=20$,
these numbers drop to a maximum of a few.

 \item X-ray detections can be substantially enhanced, if a prompt follow-up of SKA candidate
  is adopted with an instrument with flux limit  $ \lsim 10^{-11} ~\rm erg ~\rm cm^{-2} 
~\rm s^{-1}$ in the 1-10 keV band over 4-day timescale. With that flux limit each 
SKA triggered event can have in principle an X-ray counterpart (see Fig.\ref{fig:rate_x_flux_limit}). 
A suppression factor should be adopted if the X-ray emitting region would be moving with  a larger Lorentz factor.

\item The sample of SKA preselected X-ray events can extend up to redshift  $\sim 2$ for a 
X-ray instrument such as Athena and the LAD on board of a LOFT-like mission.
Instead, eRosita, the WFM on LOFT and EP samples will probe a redshift range only up to $z \lsim 1$.

\item Despite the several advantages of a radio trigger, direct X-ray detections are the only way to study the 
early stages ($< 10$ day) of the flare (see Figure \ref{fig:triggers}).

\end{itemize}

Once TDE samples in different bands have been built up, the synergy between radio, X-rays and optical
can in principle constrain important physical quantities such as the jet
luminosity, bulk Lorentz factor, the jet production efficiency and the black hole mass function.
These findings will inform theories of jet and disc formation from sudden accretion events and, on
the other hand, of SMBH cosmological evolution.

\acknowledgements
We would like to thank Francesco Shankar for providing us with the black hole mass functions, I. Prandoni and
H. Krimm for the interesting discussions about SKA and BAT observing strategies, respectively.
Finally, we thank S. Van Velzen for the careful reading of this paper and his helpful comments. 
We thank the anonymous referee for  her/his helpful suggestions. 
This work made use of data supplied by the UK Swift Science Data Centre at the University of Leicester.
 I. Donnarumma is grateful for support by ASI (under contract I/021/12/0-186/12), INAF 
(under contract PRIN-INAF-2011 ``Strong Gravity").



\label{lastpage}
\end{document}